\documentclass[12pt]{iopart}

\usepackage{graphicx,amssymb,color}

\def\be{\begin{equation}}
\def\ee{\end{equation}}
\def\pcs{\ {\rm ph}/({\rm cm}^2\ {\rm s})}

\begin{document}


\title[Jeltema \& Profumo: Gamma Rays and Dark Matter in the Galactic Center]{\mbox{Fitting the Gamma-Ray Spectrum from Dark Matter}\\[0.3cm] \mbox{with DMFIT: GLAST and the Galactic Center Region}\\[0.8cm]}

\author{Tesla E.~Jeltema}
\address{Morrison Fellow, UCO/Lick Observatories, Santa Cruz, CA 95064, USA}
\ead{tesla@ucolick.org}

\author{Stefano Profumo}
\address{Santa Cruz Institute for Particle Physics and Department of Physics,\\ University of California, Santa Cruz, CA 95064, USA}
\ead{profumo@scipp.ucsc.edu}

\begin{abstract}
We study the potential of GLAST to unveil particle dark matter properties with gamma-ray observations of the Galactic center region. We present full GLAST simulations including all gamma-ray sources known to date in a region of 4 degrees around the Galactic center, in addition to the diffuse gamma-ray background and to the dark matter signal. We introduce DMFIT, a tool that allows one to fit gamma-ray emission from pair-annihilation of generic particle dark matter models and to extract information on the mass, normalization and annihilation branching ratios into Standard Model final states. We assess the impact and systematic effects of background modeling and theoretical priors on the reconstruction of dark matter particle properties. Our detailed simulations demonstrate that for some well motivated supersymmetric dark matter setups with one year of GLAST data it will be possible not only to significantly detect a dark matter signal over background, but also to estimate the dark matter mass and its dominant pair-annihilation mode.
\end{abstract}


\maketitle


\section{Introduction}

The Gamma-ray Large Area Space Telescope (GLAST) \cite{glastref} was successfully launched on June 11, 2008. The main instrument onboard GLAST, the Large Area Telescope (LAT) represents an improvement by more than one order of magnitude over the sensitivity of its predecessor EGRET \cite{egretref} in the energy range from 20 MeV to 10 GeV, and extends the high energy coverage to about 300 GeV. These features make the LAT a tremendous tool for the indirect search for particle dark matter \cite{Baltz:2008wd}, expected in the best motivated particle models to be a weakly interacting massive particle (WIMP) with a mass in the 10-1000 GeV range \cite{dmreviews,kkdm}. WIMPs occasionally pair-annihilate in dark matter halos, producing, as a result, Standard Model particles. Prompt production as well as decays, hadronization and radiative processes associated to the annihilation products give rise to stable species including gamma rays, in an energy range extending up to the WIMP mass. Secondary radiation from the energetic electrons and positrons produced in annihilation events can also be used as an indirect detection diagnostic, e.g. in X-ray and radio wavelengths \cite{multiw}.

The signal from dark matter annihilation is proportional to the product of the particle pair-annihilation rate and to the particle number density squared along the line of sight. The latter quantity provides a guideline as to which regions of the gamma-ray sky are the most promising for the detection of a signal from dark matter annihilation. As discussed e.g. in \cite{Baltz:2008wd}, targets include the center of our own Galaxy, satellite galaxies of the Milky Way, nearby massive galaxies and clusters, as well as diffuse emission from the entire Galactic halo and the summed signal from unresolved sources at all redshifts in the extra-galactic diffuse gamma-ray flux. Numerous studies have addressed in detail all of these possibilities (for reviews, see e.g. \cite{dmreviews,kkdm}). 

With the dawn of the GLAST era, theoretical speculations on indirect WIMP detection with gamma rays will soon face high-quality data and the challenge of the corresponding analysis. The main purpose of the present study is to introduce DMFIT, a numerical package that, interfaced with any spectral fitting routine, allows one to fit gamma-ray data with the expected emission from the annihilation of generic WIMP models. For the purpose of illustration, we use here DMFIT in conjunction with XSPEC \cite{xspec}. We show fits to gamma-ray spectra simulated using the experimental response function of the LAT and the software analysis tools that will be used for the actual GLAST data analysis. We apply DMFIT to the case of an ``isolated'' gamma-ray source, such as might be associated to dark matter annihilation in Galactic dark matter clumps or local satellite galaxies, as well as to the complex case of the Galactic center region, where background modeling and an accurate understanding of conventional gamma-ray sources play a critical role.

The Galactic center region has long been considered a promising target to look for a signature from particle dark matter annihilation in gamma rays. First estimates of the detectability of a dark matter pair-annihilation signal from the center of the Galaxy date back 30 years \cite{Gunn:1978gr,Stecker:1978du}. An incomplete list of recent references that discussed the detection of dark matter pair-annihilation in the Galactic center with gamma-ray observations includes \cite{Cesarini:2003nr,Stoehr:2003hf,Aloisio:2004hy,Bloom:2004aq,Pieri:2005vp,Mambrini:2005vk,Profumo:2005xd,Jacholkowska:2005nz,Hall:2006na,Zaharijas:2006qb,Bergstrom:2006ny,Aharonian:2006wh,Ahn:2007ty,Morselli:2007ch,Hooper:2007gi,Dodelson:2007gd,Regis:2008ij,Serpico:2008ga,Baltz:2008wd}. In particular, ref.~\cite{Cesarini:2003nr} discussed the dark matter interpretation of the gamma-ray excess reported by EGRET from the Galactic center \cite{MayerHasselwander:1998hg}; ref.~\cite{Profumo:2005xd} and \cite{Aharonian:2006wh} discussed a similar possibility for the high energy gamma-ray flux detected by HESS \cite{Aharonian:2004wa}; ref.~\cite{Zaharijas:2006qb,Dodelson:2007gd} presented studies where both the EGRET and the HESS data were simultaneously taken into account as a background to dark matter searches.  In addition, recently, the GLAST collaboration gave in ref.~\cite{Baltz:2008wd} a comprehensive and updated overview of the GLAST sensitivity to dark matter annihilation signals for several possible sources inside and outside the Galaxy, using the Collaboration's current state of the art Monte Carlo and event reconstruction software.

As far as the Galactic center region is concerned, in the present study we include all gamma-ray sources known to-date within an angle of 4 degrees from the Galactic center, in addition to the diffuse Galactic gamma-ray emission. Of special importance is the question of how to properly model the innermost sources, associated to the Sag A${^*}$ region: we model those with three different scenarios, described in detail in sec.~\ref{sec:saga}. We introduce three reference particle dark matter setups, which are particularly illustrative for the scope of the present analysis, chosen to be theoretically well motivated, phenomenologically viable and producing the correct thermal relic neutralino abundance. We carry out complete one year GLAST simulations, making use of an up-to-date LAT instrumental response function and pointing mode setup, as well as of the software analysis tools that will be used to analyze the actual GLAST data. 

The main scopes of the present study are to:
\begin{enumerate}
\item present the DMFIT tool, and show examples of its application;
\item provide an updated template for the gamma-ray sources potentially relevant to dark matter searches in the Galactic center region;
\item assess the capabilities of GLAST to provide information on particle dark matter properties such as the mass and the dominant annihilation mode, both in virtually ``background-free'' setups and in the complex gamma-ray environment of the Galactic center;
\item study the theoretical bias that background modeling and theoretical priors (such as the dominant annihilation mode) produce in the estimation of the fundamental particle properties of dark matter from gamma-ray data.
\end{enumerate}

The paper is organized as follows: sec.~\ref{sec:dmfit} introduces the DMFIT tool and \ref{sec:glastsim} provides details on the GLAST simulations; sec.~\ref{sec:sources} discusses the gamma-ray sources included in the simulations, and \ref{sec:dm} describes the particle dark matter models; sec.~\ref{sec:dmfitdmonlly} shows examples of applications of DMFIT to gamma-ray spectra produced by dark matter annihilation only.  Finally, sec.~\ref{sec:gc} addresses the Galactic center region: we discuss there the optimal energy and angular regions for GLAST observations and the performance of GLAST at inferring particle dark matter properties from the gamma-ray spectrum from the center of the Galaxy.  Our summary and conclusions are given in sec.~\ref{sec:concl}.

\section{Methodology}\label{sec:method}

In this section we introduce the DMFIT tool (sec.~\ref{sec:dmfit}) and give details on the GLAST simulations we employed in the present analysis (sec.~\ref{sec:glastsim}).

\subsection{The DMFIT Tool}\label{sec:dmfit}
\begin{figure}[t]
\begin{center}
\includegraphics[width=11cm,clip]{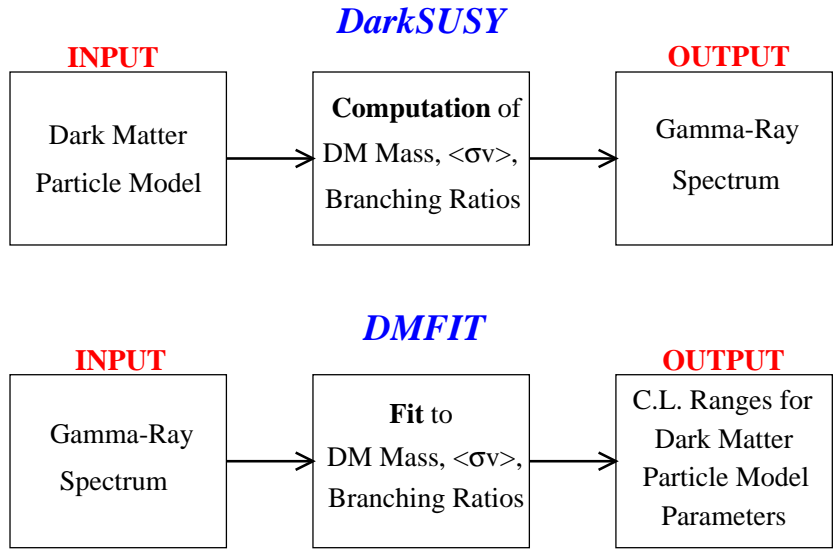}\\
\caption{A conceptual flow-chart for DarkSUSY (upper) and DMFIT (lower). \label{fig:dmfit}}
\end{center}
\end{figure}
DMFIT is a tool that calculates the gamma-ray flux resulting from the pair annihilation of generic WIMPs (i.e. of dark matter particles with specified mass and branching ratios into Standard Model final state annihilation modes). DMFIT is based on the same set of Monte Carlo simulations of hadronization and/or decay of the annihilation products used in DarkSUSY \cite{ds}. The simulations were carried out by the DarkSUSY team using {\tt Pythia} 6.154 \cite{pythia} for a set of 18 neutralino masses ranging from 10 up to 5000 GeV, and for 8 ``fundamental'' Standard Model final states, namely the quark-antiquark pairs $c\bar c$, $b\bar b$, $t\bar t$, the charged lepton pairs $\mu^+\mu^-$ and $\tau^+\tau^-$, the gauge boson pairs $W^+W^-$ and $ZZ$, and gluon pairs, $gg$. Two data files contain simulation results on the differential gamma-ray yield at given energies, and the same yield integrated above a given energy threshold. The simulation results are then interpolated (the type of interpolation can be decided by the user) for a user-supplied value of the dark matter mass and of the annihilation final state. In addition, we also include the $e^+e^-$ channel, where gamma-rays are radiated in the final state. The $e^+e^-$ channel is particularly relevant for various WIMP models, including the Kaluza-Klein dark matter of Universal Extra Dimensions \cite{kkdm}. This channel is not currently available in the latest publicly available DarkSUSY version. For it, we use the analytical approximation to the differential photon multiplicity for $\chi\chi\to e^+e^-\gamma$ provided in ref.~\cite{Bergstrom:2004nr}, namely
\begin{equation}
\frac{{\rm d}N}{{\rm d} x}\simeq\frac{\alpha}{\pi}\frac{x^2-2x+2}{x}\ln\left(\frac{m_\chi^2}{m_e^2}(1-x)\right),\quad {\rm where}\quad  x=\frac{E}{m_\chi}.
\end{equation}
While the Monte Carlo simulations extend down to a WIMP mass of 10 GeV, DMFIT allows to extrapolate to lower masses. Very light WIMPs have been recently shown to be relevant even in the context of supersymmetry \cite{Profumo:2008yg}, and they can possibly play a role in explaining the puzzling DAMA/LIBRA signal \cite{dama}. For channels with heavy particles in the final state, such as $W^+W^-$, $ZZ$ and $t\bar t$, when the WIMP mass is below the kinematic threshold given by the final state particle mass the current version of DMFIT automatically switches to the default channel $b\bar b$.

DMFIT consists of two data files and one Fortran routine. The code is available from the authors upon request. Conceptually, DMFIT reverse-engineers the use of the DarkSUSY package for the computation of gamma-ray spectra (see Fig.~\ref{fig:dmfit}). In DarkSUSY, the user supplies a given supersymmetric dark matter model, and the package computes (among other things) the lightest neutralino mass, its pair annihilation cross section and its branching ratios into Standard Model final states. The gamma-ray spectra resulting from each annihilation mode at the given neutralino mass are then summed over with a weight given by their corresponding branching ratios. The final gamma-ray spectrum resulting from the neutralino pair annihilation consists of the resulting linear combination. DMFIT, on the other hand, computes for a given single or multiple annihilation mode, the resulting gamma-ray spectrum, and can be easily interfaced with any spectral fitting package. As a result, given an input gamma-ray spectrum, DMFIT allows one to fit for the particle dark matter mass, its pair-annihilation rate and its branching ratios. In conjunction with virtually any fitting package, DMFIT can be used to reconstruct confidence level ranges for the mentioned particle dark matter properties.

For the present paper, we interfaced DMFIT with the spectral fitting package XSPEC \cite{xspec}. XSPEC allows the user to fit for a combination of more than one model at once, freezing or fitting model parameters as desired. By including more than one DMFIT model and imposing that the dark matter masses be the same, one can easily find best fit values for branching ratios into multiple final states. The XSPEC implementation of DMFIT allows one to compute confidence level contours for various particle dark matter quantities, including the particle mass, a normalization (related to the pair-annihilation cross section and to the number density of dark matter) and the relative contributions from different Standard Model final states. In the present study, we show several examples of the application of DMFIT to the reconstruction of particle dark matter properties. In the near future, we plan to add DMFIT to the models publicly available for XSPEC \cite{xspec}, and to create versions compatible with other spectral fitting packages, including those provided with the GLAST Science Tools \cite{sciencetools}.

\subsection{GLAST Simulation Setup}\label{sec:glastsim}

To simulate GLAST observations, we employ the GLAST observation simulator tool, {\tt gtobssim}, part of the GLAST Science Tools package (v9r5p2) \cite{sciencetools}.  All simulations were run for one year for a default scanning mode observation and using the Pass 5 source instrument response functions (P5\_v13\_0\_source).  For each source, spectral data files were provided to {\tt gtobssim} (see sec.~\ref{sec:srcs} for model definitions) to define the source spectrum.  For dark matter sources, images defining the source spatial distribution were also input, according to the relevant line of sight integrals of the adopted dark matter density profile squared; all other sources were assumed to be point-like.  The gamma-ray emission lines from dark matter annihilation (the $\gamma \gamma$ and $Z \gamma$ final states) were modeled separately as monochromatic sources, but we found them to be too faint to be significant in one year of data.  We include the Galactic diffuse emission using the GALPROP \cite{galprop} (v49\_600202RB) model provided as part of the GLAST external libraries distribution.  In practice, the precise model of the Galactic diffuse emission used has little effect on our results as in all of the cases we consider the background is either dominated by point source emission, or the dark matter source is bright compared to the Galactic diffuse.  We confirmed the relative insensitivity of our results to the Galactic diffuse model by also simulating the ``optimized'' GALPROP and the ``conventional'' diffuse models (see ref.~\cite{Baltz:2008wd} for a discussion on these Galactic diffuse setups). We comment on this at the end of sec.~\ref{sec:dmprop2}. We did not include the extra-galactic diffuse background, although we did simulate it: in the Galactic center region, the extra-galactic diffuse background is irrelevant.  Employing the power-law parametrization resulting from the analysis of the EGRET data \cite{egreteg} (which was been shown to likely be an overestimate of the actual extra-galactic gamma-ray flux in \cite{extragal}), this component contributes, in an angular region of $1^\circ$ around the center of the Galaxy, only 1-2\% of the diffuse Galactic flux and around 50 photon counts above 1 GeV in one year of observation. We discuss this in more detail in sec.~\ref{sec:angle}.

\section{Gamma-Ray Sources in the Galactic Center Region}\label{sec:srcs}

\subsection{Astrophysical Sources}\label{sec:sources}

We include in the present study all gamma-ray sources detected to-date in an angular region of 4 degrees around the Galactic center. We model each source according to either fits to available gamma-ray data, or to spectral models resulting from assumptions on the nature of the source, derived e.g. from multi-wavelength observations. We describe our models for each gamma-ray source below; the list is in order of decreasing angular distance from the Galactic center. We conclude with a summary in Tab.~\ref{tab:SRC} of the source locations and integrated gamma-ray fluxes above 0.1, 1 and 5 GeV 

\subsubsection{3EG J1736-2908}

Originally classified as unidentified \cite{3EG}, after INTEGRAL observations this EGRET source was identified with the X-ray source GRS1734-292 \cite{DiCocco:2004jd}, associated with the active Galactic nucleus of a Seyfert 1 galaxy at a redshift of 0.0214 and 1.8 degrees from the Galactic center, having both radio jet and hard X-ray emissions \cite{Marti:1997sp,Sazonov:2004ne,Sazonov:2004pd}. The analysis of ref.~\cite{Nolan:2003bt} indicates that the EGRET source 3EG J1736-2908 exhibits significant time variability; Here, we consider the source median emission \cite{3EG}. 3EG J1736-2908 has no positional counterpart at TeV energies in the HESS survey of the inner Galaxy \cite{Aharonian:2005kn}, which leads to an upper limit on the source above $\sim 100$ GeV. Ref.~\cite{reimerbertsch} showed that the best fit to 3EG J1736-2908 consists of a broken power law. Here we follow the analysis of ref.~\cite{Funk:2007zzd}, and adopt for 3EG J1736-2908 the following spectrum:
\begin{equation}
\hspace*{-2cm}\frac{{\rm d}N}{{\rm d}E}=\frac{7.5\times 10^{-11}}{{\rm cm}^2\ {\rm s}\ {\rm MeV}}\left(\frac{E}{1\ {\rm GeV}}\right)^{-\lambda},\qquad \lambda=1.44\ (E<1\ {\rm GeV}),\ \lambda=5.7\ (E>1\ {\rm GeV}).
\end{equation}
The integrated flux above 0.1 GeV for this spectral model is $3.15\times 10^{-7}$ photons per cm${}^2$ per s. The integrated flux above 200 GeV is $2.4\times 10^{-19}\pcs$, thus fully compatible with HESS limits \cite{Aharonian:2005kn}. The location of the source is assumed to be coincident with the location of GRS1734-292 \cite{Marti:1997sp}.

\subsubsection{3EG J1744-3011, HESS J1745-303}
\begin{figure}[t]
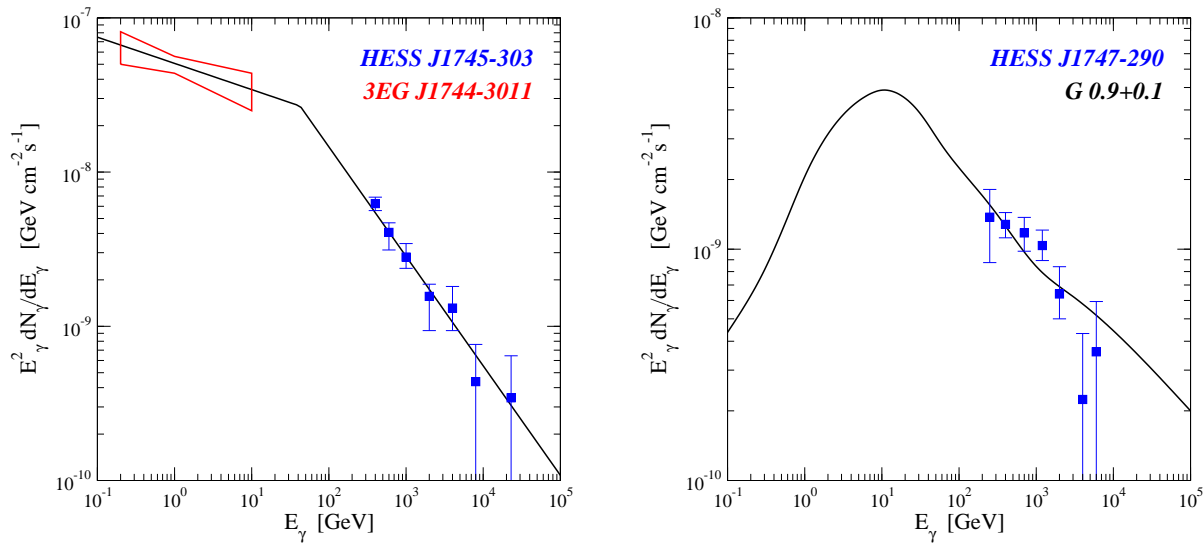

\begin{center}
\mbox{\includegraphics[width=7.5cm,clip]{FIGURES/EGRB_SED.eps}\qquad \includegraphics[width=7.5cm,clip]{FIGURES/SNR_SED.eps}}\\
\caption{Left: SED model for the EGRET unidentified source 3EG J1744-3011 \cite{3EG}, assumed to be coincident with the TeV gamma-ray source HESS J1745-303 \cite{Aharonian:2008gw}. Right: SED model for G 0.9+0.1, and HESS data \cite{Aharonian:2005br}. \label{fig:astrosources}}
\end{center}
\end{figure}
Source counterparts for the extended unidentified very high energy gamma-ray source HESS J1745-303, 1.4 degrees from the Galactic center, were recently examined in ref.~\cite{Aharonian:2008gw}. Among possible matches, ref.~\cite{Aharonian:2008gw} discusses a supernova-remnant/molecular cloud association and a high spin-down-flux pulsar. The unidentified EGRET source 3EG J1744-3011 \cite{3EG} is also a plausible association from an energetic standpoint \cite{Funk:2007zzd}, while the positional coincidence with the HESS source is not conclusive \cite{Aharonian:2008gw,Aharonian:2005br}. However, the position of HESS J1745-303 is well within the 95\% uncertainty level region for 3EG J1744-3011. The third EGRET catalog quotes for 3EG J1744-3011 an integrated photon flux of $(63.9\pm7.1)\times 10^{-8} \pcs$ and a best-fit value for the spectral index in the 0.1 to 10 GeV range of $\Gamma=2.17\pm0.08$. The HESS data \cite{Aharonian:2008gw} indicate for HESS J1745-303 a spectral index $\Gamma=2.71\pm0.11_{\rm stat}\pm0.2_{\rm sys}$ (significantly softer than what was originally reported in \cite{Aharonian:2005kn}), and an integral flux between 1 and 10 TeV of $(1.63\pm0.16)\times 10^{-12}\pcs$. Here, we assume that (a) HESS J1745-303 is a point-like source, (b) 3EG J1744-3011 is the same source (hence has the same position) as HESS J1745-303, and (c) the following spectrum:
\begin{eqnarray}
& \frac{{\rm d}N}{{\rm d}E}=\frac{1.64\times 10^{-4}}{{\rm cm}^2\ {\rm s}\ {\rm MeV}}\left(\frac{E}{1\ {\rm MeV}}\right)^{-2.17}& \ {\rm for}\ E<42.3\ {\rm GeV}\\
& \frac{{\rm d}N}{{\rm d}E}=\frac{5.17\times 10^{-2}}{{\rm cm}^2\ {\rm s}\ {\rm MeV}}\left(\frac{E}{1\ {\rm MeV}}\right)^{-2.71}& \ {\rm for}\ E>42.3\ {\rm GeV}
\end{eqnarray}
We determined the location of the power-law break, $E_b=42.3$ GeV, by requiring a match to the integrated photon fluxes individually quoted for 3EG J1744-3011 \cite{3EG} and for HESS J1745-303 \cite{Aharonian:2008gw}. We show the EGRET bow-tie, the HESS data and the spectrum outlined above in the left panel of Fig.~\ref{fig:astrosources}. As discussed in ref.~\cite{Aharonian:2008gw}, the gamma-ray flux from hadronic sources is generically proportional to $E^{-(\gamma_p+\delta)}$, where $\gamma_p$ is the proton index at the source and $0.3\lesssim\delta\lesssim0.6$ is the index of the diffusion coefficient \cite{aaa}. This allows for spectra that are quite soft in the TeV regime, and with different slopes in other energy bands \cite{Torres:2002af}. This argument motivates the broken power-law spectrum assumed here (and shown in the left panel of Fig.~\ref{fig:astrosources}), although a more complex setup is not excluded (see e.g. the discussion in \cite{Aharonian:2008gw}). Using the spectral model outlined above, we obtain a total flux above 0.1 GeV of $6.4\times 10^{-7}\pcs$.

\subsubsection{HESS J1747-281 [G 0.9+0.1]}
Very high energy gamma-rays were detected in 2004 by the HESS instrument from the composite supernova remnant G 0.9+0.1, approximately 1 degree from the Galactic center \cite{snr}, and reported in ref.~\cite{Aharonian:2005br}. The source is one of the weakest TeV sources ever detected, and is not associated with any counterpart in the EGRET catalogs \cite{3EG}. The location of the source is consistent within the statistical errors with the position of the pulsar wind nebula in G 0.9+0.1 \cite{Gaensler:2001gx}. Ref.~\cite{Aharonian:2005br} estimates an integrated flux above 200 GeV of $(5.7\pm0.7_{\rm stat}\pm1.2_{\rm sys})\times 10^{-12}\pcs$, assuming a photon index $\Gamma=2.40\pm0.11_{\rm stat}\pm0.2_{\rm sys}$. The broadband emission from G 0.9+0.1 was fitted in ref.~\cite{Aharonian:2005br} with a one-zone inverse Compton model featuring a parent population of accelerated electrons with a broken power-law spectrum (spectral index 0.6 below 25 GeV and of 2.9 above 25 GeV), and assuming a uniform magnetic field strength $B=6\ \mu$G within the pulsar wind nebula. The dominant radiation field off of which the accelerated electrons inverse Compton scatter is starlight with an energy density of 5.7 eV/cm${^3}$ \cite{Aharonian:2005br}. Here, we assume the same setup, and show the spectral energy distribution ($E^2\  {\rm d}N/{\rm d}E$) we obtain in the right panel of Fig.~\ref{fig:astrosources}, together with the HESS data. The spectral model we adopt gives an integrated flux above 0.1 GeV of $1.05\times 10^{-8}\pcs$ and of $5.5\times 10^{-12}\pcs$ above 200 GeV, and is consistent with the EGRET non-detection of this source. 

\subsubsection{3EG J1746-2851 and HESS J1745-290 [Sgr A${}^*$]}\label{sec:saga}
\begin{figure}[t]
\begin{center}
\includegraphics[width=14.5cm,clip]{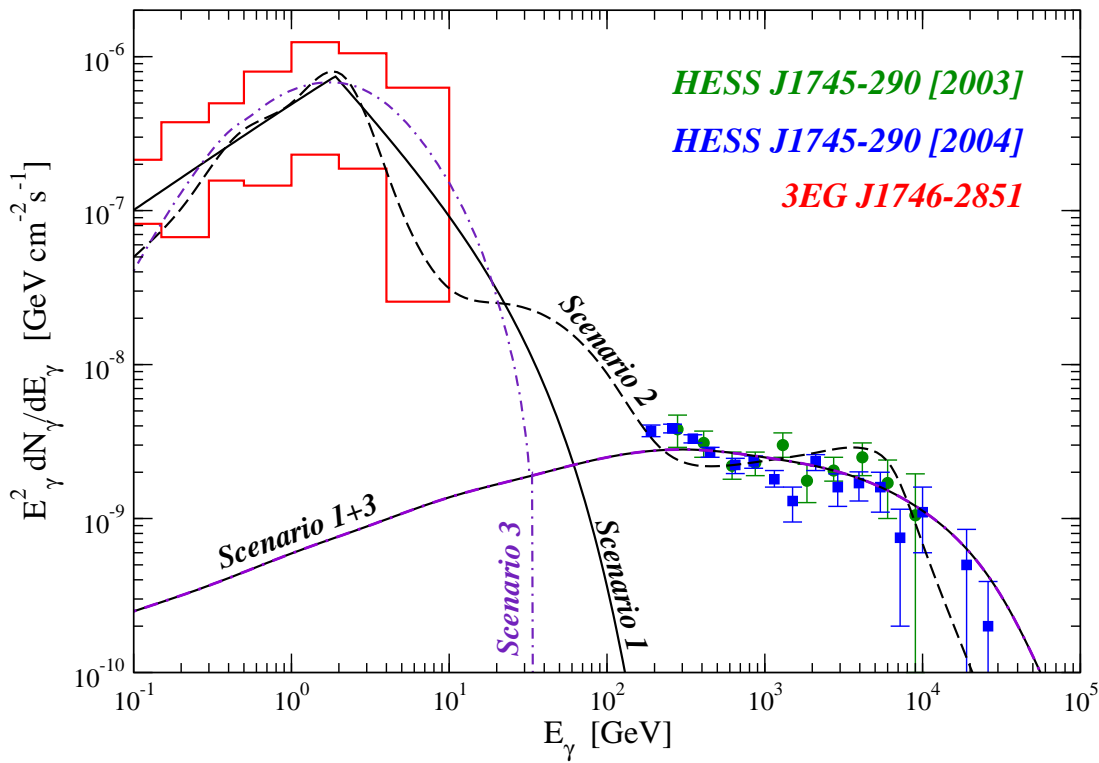}\\
\caption{The three scenarios for the broadband SED from the innermost part of the Galactic center region. The red box indicates the EGRET data on the unidentified source 3EG J-1746-2851 \cite{3EG,MayerHasselwander:1998hg}. The HESS data refer to the 2003 (green) \cite{Aharonian:2004wa} and 2004 (blue) \cite{Aharonian:2006wh} observations of HESS J1745-290. \label{fig:sga}}
\end{center}
\end{figure}

EGRET observed a pronounced source excess at the Galactic center position \cite{mh}, subsequently designated in the second (third) EGRET catalog by 2EG (3EG) J1746-2852 \cite{2EG, 3EG}. The source location in the energy range above 500 MeV indicated perfect compatibility with the Galactic center \cite{Lamb:1997qs}. A subsequent re-analysis \cite{Hooper:2002ru,Hooper:2002fx} used the point spread function as determined by the pre-flight EGRET calibration \cite{2EG} for 6 energy bins above 1 GeV, and found that the location of 3EG J1746-2851 is off-set from the Galactic center at a high confidence level. Ref.~\cite{Hooper:2002fx} indicates that the best fit source position is at $l=0.19$ and $b=-0.08$. Subtracting the diffuse emission and allowing for a total source-excess extent up to $1.5^\circ$, ref.~\cite{MayerHasselwander:1998hg} attributes to 3EG J1746-2851 a flux excess of $(217\pm15)\times 10^{-8}\pcs$ above 0.1 GeV. The photon spectrum quoted in \cite{MayerHasselwander:1998hg} is well represented by a broken power law with a break energy of 1.9 GeV. The best fit broken power-law spectrum from the EGRET data is
\begin{eqnarray}\label{eq:egreta}
&\frac{{\rm d}N}{{\rm d}E}=\frac{2.2\times 10^{-10}}{{\rm cm}^2\ {\rm s}\ {\rm MeV}}\left(\frac{E}{1900\ {\rm MeV}}\right)^{-\lambda},\quad {\rm with}&\\[0.4cm]
\nonumber&\lambda=1.30\pm0.03\ (E<1.9\ {\rm GeV})\ {\rm and}\ \lambda=3.1\pm0.2\ (E>1.9\ {\rm GeV})&
\end{eqnarray}
Ref.~\cite{Cesarini:2003nr} showed that the spectrum reported in \cite{MayerHasselwander:1998hg} can also be well fitted by a scenario where, in addition to the diffuse Galactic gamma-ray background, 3EG J1746-2851 is fueled by WIMP pair annihilation in the Galactic center. This interpretation prefers rather light WIMPs ($m_\chi\sim 40$ GeV) and large pair annihilation cross sections, or large dark matter densities in the vicinity of Sgr A${}^*$ \cite{Cesarini:2003nr}. We show the EGRET data as the red solid contours in Fig.~\ref{fig:sga}. The somewhat conservative error bars also include the uncertainty in the energy determination according to the binning employed in \cite{MayerHasselwander:1998hg}.

Gamma-ray emission above 100 GeV from the direction of the Galactic center was recently reported by several ground-based gamma-ray observatories, including CANGAROO \cite{Tsuchiya:2004wv}, VERITAS \cite{Kosack:2004ri}, HESS \cite{Aharonian:2004wa,Aharonian:2006wh,Aharonian:2006au} and MAGIC \cite{Albert:2005kh}. Here, we will focus on the high-statistics 2003 and 2004 HESS data \cite{Aharonian:2004wa,Aharonian:2006wh} from the point-like source HESS J1745-290, compatible with the gravitational center of the Galaxy. No unique identification of HESS J1745-290 has been possible so far, but at least three different astrophysical objects have been discussed in the literature: First, several models predict high energy gamma-ray emission near the super-massive black hole Sgr A${}^*$ (see e.g. \cite{Aharonian:2004jr}); Second, the location of HESS J1745-290 is compatible\footnote{See however Ref.~\cite{hesssage}, where it was shown that after reducing systematic pointing errors and analyzing the 2005/2006 data Sgr A East is ruled out as the counterpart of the HESS source.} with the supernova remnant Sgr A East, featuring bright shell-like radio emission surrounding Sgr A${}^*$ itself \cite{Maeda:2001kv}; Third, a candidate pulsar wind nebula, G359.95-0.04, was recently discovered $8.7^{\prime\prime}$ away from Sgr A${}^*$  in a deep Chandra survey of the Galactic center region \cite{Wang:2005ya}. In addition, the possibility of associating HESS J1745-290 with WIMP dark matter annihilation was addressed in \cite{Profumo:2005xd,Aharonian:2006wh}. The latter interpretation would require large WIMP masses and pair annihilation cross sections, which, although theoretically possible, appear to be rather unnatural from a theoretical particle physics standpoint \cite{Profumo:2005xd}. Fig.~\ref{fig:sga} shows in green the HESS data from the 2003 observations \cite{Aharonian:2004wa} and in blue those from the 2004 observations \cite{Aharonian:2006wh}.

In the present study we consider three different scenarios to model the gamma-ray sources 3EG J1746-2851 and HESS J1745-290:
\begin{itemize}
\item {\bf Scenario 1}. The two gamma-ray sources are two different individual sources, with 3EG J1746-2851 offset from the Galactic center as in \cite{Hooper:2002ru,Hooper:2002fx}. A two-source model was for instance proposed in ref.~\cite{Crocker:2004bb}, where HESS J1745-290 is associated to Sgr A${}^*$, while 3EG J1746-2851 is mostly fueled by the supernova remnant Sgr A East. Extrapolating Eq.~(\ref{eq:egreta}) up to energies probed by HESS, however, vastly over-predicts the flux of very high energy gamma-rays actually measured. Several mechanisms can explain an effective cutoff in the spectrum of 3EG J1746-2851 at energies larger than 10 GeV \cite{Crocker:2004bb}, including the possibility that the EGRET source is associated to a young pulsar with gamma-ray properties similar to Vela, but with a larger gamma-ray power \cite{Atoyan:2004ix}. To model the EGRET source, we modify here the 3EG J1746-2851 spectrum from \cite{MayerHasselwander:1998hg} multiplying the spectrum in (\ref{eq:egreta}) by an exponential cut-off factor $\exp(-E/E_c)$. As in \cite{Dodelson:2007gd}, we choose the cutoff scale $E_c=30$ GeV, which could indeed be plausible in the young pulsar scenario \cite{Atoyan:2004ix}. The model is shown by the black solid line in Fig.~\ref{fig:sga} labeled ``Scenario 1''. The integrated photon flux above 0.1 GeV for this source is $2.1\times 10^{-6}\pcs$.

As far as the HESS J1745-290 source is concerned, we adopt here the black hole plerion model of ref.~\cite{Crocker:2004bb}. In this scenario, a sub-relativistic outflow of particles from an inner, inefficiently radiating magnetized corona, i.e. the advection-dominated accretion flow, powers a black-hole plerion where both the X-ray and TeV gamma-ray emissions are produced by electrons accelerated at the wind shock. This setup has the virtue of explaining several features of the broadband emission from Sgr A${}^*$. We show the spectral energy distribution resulting from this model with the solid line labeled ``Scenario 1+3'', since the same model for the TeV emission is employed in Scenario 3. Models involving a hadronic origin for  HESS J1745-290 are also not excluded. According to the results of ref.~\cite{Aharonian:2004jr}, the broadband emission of several hadronic models, extrapolated in the energy range relevant for GLAST, would be comparable to what we use here. In particular, we implemented models based on both photo-meson processes and on proton-proton collisions, as described in \cite{Aharonian:2004jr}, and find that the impact on the ability of GLAST to reconstruct dark matter particle properties using these models instead of the black hole plerion setup is negligible. To appreciate this point, we compare the integrated photon flux above 0.1 GeV for the three mentioned models. We obtain a flux of $4.2\times 10^{-9}\pcs$ for the black hole plerion model adopted here, of $5.6\times 10^{-9}\pcs$ for the model based on photo-meson processes, and of $2.3\times 10^{-8}\pcs$ for the model with gamma radiation from proton-proton interactions in the accretion disk. These figures must be contrasted with the much larger flux associated with the EGRET source, namely $2.1\times 10^{-6}\pcs$.

\item {\bf Scenario 2}. We assume for this scenario that 3EG J1746-2851 and HESS J1745-290 actually correspond to the same source, and are thus positionally coincident with the Galactic center. We refer to the spectrum resulting from the curvature radiation-inverse Compton model described in \cite{Aharonian:2004jr}. In this scenario, electron acceleration is produced by the ordered rotation-induced electric fields near Sgr A${}^*$ \cite{neronov}. Electron radiative losses consist of both curvature radiation and inverse Compton scattering. While the inverse Compton scattering on IR photons of the highest energy electrons produces an emission peaking around 100 TeV, the curvature radiation peaks at significantly lower energies, namely around 1 GeV in the model considered in \cite{Aharonian:2004jr}, possibly reproducing the gamma-ray emission detected by EGRET. The details of the GeV peak depend on the configuration of the magnetic field in the acceleration zone, which in turn could spoil the assumption of isotropic electron emission assumed in Fig.~6 of \cite{Aharonian:2004jr}, where the GeV emission actually exceeds the EGRET data. We therefore assumed that the broadband emission for this model is compatible with the data in  \cite{MayerHasselwander:1998hg} by assuming a suppressed curvature radiation emission in the GeV range. We show the resulting spectral energy distribution with the dashed line in Fig.~\ref{fig:sga}, labeled ``Scenario 2''. The integrated gamma-ray flux above 0.1 GeV for this scenario is $1.9\times 10^{-6}\pcs$.

\item {\bf Scenario 3}. As pointed out above, 3EG J1746-2851 might be associated with dark matter annihilation. We assume for this scenario that this is indeed the case, and define below a supersymmetric dark matter setup (DM model C) that provides a good fit to the EGRET data, while at the same time being consistent with the other requirements we impose from the particle physics side. To account for the TeV emission, we augment this scenario with the same black hole plerion model employed for Scenario 1. Fig.~\ref{fig:sga} shows both the gamma-rays from dark matter (violet dot-dashed curve, labeled ``Scenario 3'') and indicates that we assume the same TeV emission spectrum extrapolation for Scenarios 1 and 3.

\end{itemize}

We summarize the positions and integrated gamma-ray fluxes (above 0.1, 1 and 5 GeV) for all the sources considered in the present study in Tab.~\ref{tab:SRC}.

\begin{table}
\caption{Summary of the positions and integrated gamma-ray fluxes (above 0.1, 1 and 5 GeV) from the gamma-ray sources in an angular region of 4 degrees around the Galactic center. Units for the photon fluxes are $10^{-8}$ photons per cm${}^2$ per s.\label{tab:SRC}}
\begin{indented}
\item[]\begin{tabular}{cccccc}
\br
 & l & b &  0.1 GeV & 1 GeV & 5 GeV \\
\mr
3EG J1736-2908 & 358.9 & 1.4 & 31.5 & 1.6 & $8\times 10^{-4}$\\
3EG J1744-3011 & 358.7 & -0.64 & 64.0 & 4.3 & 0.64 \\
HESS J1747-281 & 0.87 & 0.077 & 1.10 & 0.34  & 0.09 \\
3EG J1746-2851 & 0.19 & -0.08 & 212 & 46 & 1.95 \\
HESS J1745-290 & 359.9 & 0.03 & 0.42 & 0.10 & 0.03 \\
Sgr A${}^*$ - Sc.2 & 359.9 & 0.03& 189 & 51 & 0.87\\
\br
\end{tabular}
\end{indented}
\end{table}

\subsection{Dark Matter Models}\label{sec:dm}
Our choice of the particle dark matter models for the present study was motivated by the following four guidelines:
\begin{enumerate}
\item we wish to span a reasonable range of masses and final state branching ratios;
\item we choose theoretically well motivated particle physics frameworks that can be easily reproduced with publicly available computational tools;
\item we require a neutralino thermal relic abundance in accord with the cosmological cold dark matter density \cite{Komatsu:2008hk};
\item we require models be consistent with gauge coupling unification as well as with collider searches and other particle physics constraints.
\end{enumerate}

\begin{table}
\caption{The input setup for DM models A and B; Units are GeV.\label{tab:DM_MODAB}}
\begin{indented}
\item[]\begin{tabular}{cccccc}
\br
 & $m_0$ & $M_{1/2}$ & $A_0$ & $\tan\beta$ & $\mu$\\
\mr
{\bf A} & 650 & 420 & 1460 & 49.5 & $>0$  \\
{\bf B} & 1910 & 267 & 0 & 40 & $>0$ \\
\br
\end{tabular}
\end{indented}
\end{table}

\begin{table}
\caption{The input setup for DM model C; Units are GeV, and all quantities are defined at the weak scale.\label{tab:DM_MODC} }
\begin{indented}
\item[]\begin{tabular}{ccccccccc}
\br
 & $M_1$ & $M_2$ & $\mu$ & $m_A$ & $\tan\beta$ & $m_{\widetilde{Q}}$ & $A_t$ & $m_{\widetilde{L}}$\\
\mr
{\bf C} & 33.5 & 200 & -200 & 105 & 10 & 300 & 450 & 500 \\
\br
\end{tabular}
\end{indented}
\end{table}
Our models A and B are defined, in the context of the constrained minimal supersymmetric Standard Model (CMSSM), by the Grand Unification scale values of the universal scalar soft-breaking mass $m_0$, gaugino mass $M_{1/2}$, trilinear scalar coupling $A_0$, by the ratio of the two Higgs doublets vacuum expectation values ($\tan\beta$) and by the sign of the supersymmetric higgsino mass term $\mu$. We specify all of these parameters in Tab.~\ref{tab:DM_MODAB}. Renormalization group running for the GUT scale parameters is crucial in the context of the MSSM, particularly for large values such as some of those we use: we make use here of ISAJET v.7.69 \cite{isajet}.

As described above, we picked model C augmenting the four requirements above by the request that the gamma-ray spectrum give a reasonable fit to the spectrum of the unidentified source 3EG J1746-2851 close to the Galactic center. In turn, this implies a low neutralino mass, $m_\chi\lesssim 40$ GeV, as shown e.g. in \cite{Cesarini:2003nr}. Such low neutralino masses are not easily found in the context of the CMSSM, while they can naturally arise in models with a light Higgs spectrum, see e.g. \cite{lightneut,Profumo:2008yg}. We thus resorted to defining DM model C by specifying weak-scale values for the relevant supersymmetric parameters (see Tab.~\ref{tab:DM_MODC}). Notice that DM model C is very close to the benchmark scenario {\em $m_h$-max} of ref.~\cite{lephiggs}, but the input parameters were massaged in order to achieve the desired neutralino relic abundance and to suppress the branching ratio $b\to s\gamma$ through cancellations between the squark and charged Higgs mediated contributions.

\begin{table}
\caption{Snapshots of the particle spectra. Masses are in GeV.\label{tab:DM_SPEC}}
\begin{indented}
\item[]\begin{tabular}{cccccccc}
\br
 & $m_h$ & $m_{\widetilde{g}}$ &$m_{\widetilde{t}_1}$ &$m_{\widetilde{\tau}_1}$ &$m_{\widetilde{Q}}$ & $m_{\widetilde{\chi}^\pm_1}$ & $m_{\widetilde{\chi}^0_2}$\\
\mr
{\bf A} & 114 & 1005 & 806 & 182& 1071 & 325 & 324 \\
{\bf B} & 117 & 738 & 1200 & 1626 & 1966 & 136 & 144 \\
{\bf C} & 100 & 800 & 191 & 498 & 295 & 159 & 158 \\
\br
\end{tabular}
\end{indented}
\end{table}
\begin{table}
\caption{Astrophysically relevant quantities for DM models A, B and C: the lightest neutralino mass, relic abundance, pair annihilation cross section and branching ratios for pair annihilation into $b\bar b$, $\tau^+\tau^-$, $W^+W^-$, $\gamma\gamma$ and $\gamma Z$. Other final states are present, but contribute negligibly.\label{tab:DM_ASTRO}}
\begin{indented}
\item[]\hspace*{-3.cm}\begin{tabular}{ccccccccc}
\br
 & $m_{\chi}/$GeV & $\Omega_{\chi}h^2$ & $\langle\sigma v\rangle/(10^{-26}\ {\rm cm^3}/{\rm s})$ & BR($b\bar b$) &BR($\tau^+\tau^-$) & BR($W^+W^-$) & BR($\gamma\gamma$)& BR($\gamma Z$)\\
\mr
{\bf A} & 167.5 & 0.114 & 1.10 & 84.4\% & 15.2\% & 0.01\% &$1.1\times 10^{-4}$ &$1.7\times 10^{-5}$\\
{\bf B} & 92.8 & 0.100 & 1.42  & 7.2\% & 0.6\% & 90.4\% & $1.1\times 10^{-3}$&$2.3\times 10^{-4}$\\
{\bf C} & 33.8 & 0.103 & 1.86 & 92.4\% & 7.2\% & 0\% & $9.7\times 10^{-7}$ & 0\\
\br

\end{tabular}
\end{indented}
\end{table}

We give snapshots of the particle spectra of models A, B and C in Tab.~\ref{tab:DM_SPEC}. In the table we list the masses, in GeV, of the lightest CP even Higgs ($m_h$), of the gluino ($m_{\widetilde{g}}$), the lightest stop ($m_{\widetilde{t}_1}$) and stau ($m_{\widetilde{\tau}_1}$), the lightest first generation squark ($m_{\widetilde{Q}}$), the lightest chargino ($m_{\widetilde{\chi}^\pm_1}$), and the next-to-lightest neutralino ($m_{\widetilde{\chi}^0_2}$). The lightest neutralino mass is listed in the first column of Tab.~\ref{tab:DM_ASTRO}. We computed the particle spectra of models A and B with ISAJET 7.69 \cite{isajet}. As a side comment, we notice that all the models we employ feature rather light particle masses. We expect all of these models to give detectable signatures at the Large Hadron Collider (see e.g. the estimates of ref.~\cite{baertata} and \cite{cms}). 

We collect in Tab.~\ref{tab:DM_ASTRO} astrophysically relevant properties of the models under consideration here. In particular, we specify the lightest neutralino mass ($m_{\chi}$) in GeV, the neutralino relic abundance ($\Omega_{\chi}h^2$), the thermally averaged pair annihilation cross section times velocity, computed at $T=0$ ($\langle\sigma v\rangle$) in units of $10^{-26}$ cm${}^3/$s, and the branching ratios into the final states $b\bar b$, $\tau^+\tau^-$, $W^+W^-$, $\gamma\gamma$ and $\gamma Z$. Notice that both the $W^+W^-$ and the $\gamma Z$ are not kinematically allowed for DM model C. We computed the quantities shown in Tab.~\ref{tab:DM_ASTRO} with DarkSUSY \cite{ds}. 

Different mechanisms drive the neutralino relic abundance in the three models we consider here. In particular, model A lies in the so-called stau co-annihilation region \cite{coan}, where the quasi-degeneracy of the lightest neutralino and stau entails a suppression of the otherwise excessive neutralino relic abundance via efficient stau-neutralino and stau-stau annihilation processes. DM model B lies in the focus point region \cite{fp}, where the $\mu$ parameter is driven to a relatively small value by a large GUT-scale input universal scalar soft-breaking mass. A large higgsino fraction and co-annihilation processes with the next-to-lightest higgsino-like neutralinos and chargino dictate a sizable effective annihilation rate, with a dominant gauge boson final state annihilation mode. In addition, as expected, we get a large branching ratio for the monochromatic $\gamma\gamma$ annihilation mode, in excess of $10^{-3}$ \cite{ullioline}. DM model C features a light neutralino that mainly pair-annihilates through light Higgs exchange. It can be thought of as a ``bulk region'' type model \cite{baertata}.

Since the scope of the present analysis is to assess the potential of GLAST to pinpoint particle dark matter properties, we purposely choose here a somewhat optimistic dark matter density profile towards the Galactic center. The reference setup is an adiabatically contracted version of the Navarro et al. profile outlined in \cite{N03}. This model is implemented in DarkSUSY as profile ``{\tt adiabsm}'' and is close to the widely used Moore profile \cite{moore}. Our reference model features in fact an inner slope scaling with the galacto-centric radius approximately as $r^{-1.5}$. Our choice is physically well motivated by the scenario of adiabatic contraction of dark matter around the central super-massive black hole Sgr A${}^*$ \cite{adiab}. For the sake of comparing with previous studies, in addition to our reference profile we also consider a Navarro, Frenk and White (NFW) profile \cite{nfw} for DM model B, with scaling radius $r_s=20$ kpc and scaling density $\rho_s=0.3\ {\rm GeV}/{\rm cm}^3$. Other dark matter density profiles are also consistent with available data on the distribution of matter in the Galaxy \cite{connie}, including cored profiles \cite{burkert}. For the latter, supersymmetric dark matter models would generically give a negligible gamma-ray flux, and an analysis similar to what we present here would be impossible. Notice that we neglect here the effect of dark matter clumpiness, that generically contributes to enhance the gamma-ray signal from dark matter annihilation \cite{vialactea}. 

The gamma-ray signal from dark matter annihilation can be cast, in the notation of \cite{Cesarini:2003nr}, as
\begin{equation}\label{eq:flux}
\hspace*{-2.5cm}\frac{{\rm d}N}{{\rm d}E}(E,\psi,\Delta\Omega)=\frac{3.74\times 10^{-10}}{{\rm cm}^2\ {\rm s}\ {\rm GeV}}J(\psi,\Delta\Omega)\left(\frac{\langle\sigma v\rangle}{10^{26}\ {\rm cm}^3\ {\rm s}^{-1}}\right)\left(\frac{50\ {\rm GeV}}{m_\chi}\right)^2\sum_f\frac{{\rm d}N_f}{{\rm d}E}B_f,
\end{equation}
where 
\begin{equation}
J(\psi,\Delta\Omega)=\frac{1}{8.5\ {\rm kpc}}\left(\frac{1}{0.3\ {\rm GeV}\ {\rm cm}^{-3}}\right)\int_{\Delta\Omega}{\rm d}\Omega\ \int_{\psi\ {\rm l.o.s.}}\rho_{\rm DM}^2(l)\ {\rm d}l(\psi)
\end{equation}
is the line-of-sight dark matter density squared in the $\psi$ direction averaged within the solid angle $\Delta\Omega$, and where ${\rm d}N_f/{\rm dE}$ is the relative differential photon yield for Standard model final state $f$ per pair-annihilation event. Our reference profile features, in the direction $\psi=0$ of the Galactic center, $J(0,9.57\times 10^{-6})=97.7$ for solid angles corresponding to half-apertures of $0.1^\circ$, $J(0,9.57\times 10^{-4})=154.2$ for $1^\circ$ and $J(0,0.379)=207.7$ for $20^\circ$. Notice that the input flux for our numerical simulations does not include any angular average. 

In order to fit the EGRET data on 3EG J1746-2851 with our DM model C, we needed to rescale the integrated line-of-sight dark matter density squared for that model by a factor 2.3. This means that for DM model C, the profile we use is assumed to be slightly less peaked in the innermost region (a slope intermediate between a NFW and a Moore profile would easily accommodate this), or has a smaller scaling density, or both. The determination of the dark matter profile from observational data certainly allows for a factor of order one uncertainty in the integrated line-of-sight dark matter density squared.

\begin{table}
\caption{Integrated gamma-ray fluxes from neutralino pair-annihilation in given angular regions, and above the specified energy threshold. Units are $10^{-8}$ photons per cm${}^2$ per s.\label{tab:DM_GR}}
\begin{indented}
\item[]\hspace*{-3cm}\begin{tabular}{cccccccc}
\br
 & $1^\circ$, 0.1 GeV &  $1^\circ$, 1 GeV &  $1^\circ$, 5 GeV &   $20^\circ$, 0.1 GeV &  $0.1^\circ$, 0.1 GeV &$1^\circ$, $\gamma\gamma$ &  $1^\circ$, $\gamma Z$ \\
\mr
{\bf A} & 18.9 & 9.6 & 2.5 & 25.4 & 11.9 & $1.2\times 10^{-4}$ & $9.7\times10^{-6}$ \\
{\bf B} & 63.0 & 23.4 & 3.7 & 84.8 & 39.9 & $1.1\times 10^{-3}$& $2.6\times 10^{-3}$\\
{\bf C} & 208 & 53.6 & 3.26 & 280.7 & 132 & $2.0\times 10^{-5}$ & 0 \\
{\bf B}, NFW & 0.49 & 0.18 & 0.028 & 6.29 & 0.05 & $8.7\times 10^{-6}$ & $2.0\times 10^{-5}$\\
\br
\end{tabular}
\end{indented}
\end{table}
We quote in Tab.~\ref{tab:DM_GR} the integrated gamma-ray fluxes, in different angular regions and above different gamma-ray energy thresholds, for the three DM models and for model B for the case of the NFW profile. We also give the fluxes from the $\gamma\gamma$ and $\gamma Z$ final state modes. Notice that the scaling of fluxes with angle for the different energy thresholds can be trivially obtained from what shown: for instance, the $\gamma\gamma$ flux within $20^\circ$ will be given by the value shown for $1^\circ$, divided by the continuum flux above 0.1 GeV for $1^\circ$, times the same flux over $20^\circ$. Recall that the $\gamma Z$ monochromatic emission occurs at a gamma-ray energy $E_{\gamma Z}=m_\chi(1-m_Z^2/(4m_\chi^2))$ and the $\gamma \gamma$ monochromatic emission occurs at $E_{\gamma \gamma}=m_\chi$ \cite{ullioline}.

\begin{figure}[t]
\begin{center}
\includegraphics[width=8.5cm,clip]{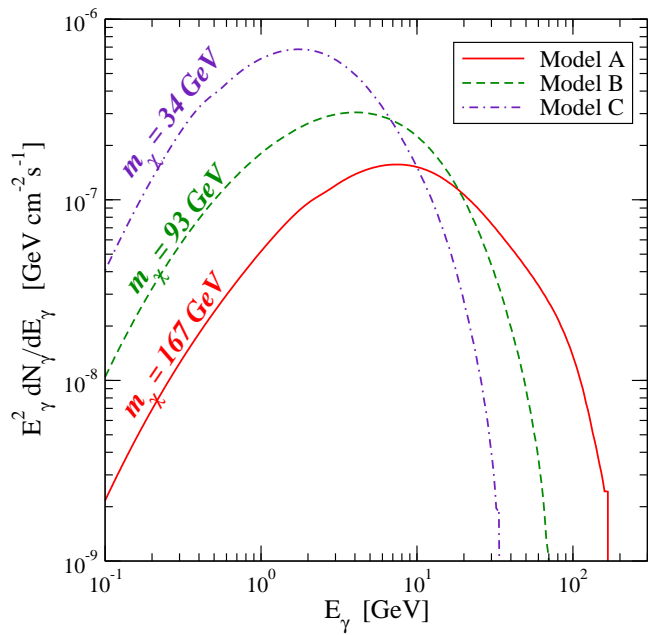}\\
\caption{SED for the three dark matter models under consideration.\label{fig:dmsources}}
\end{center}
\end{figure}
We show the spectral energy distribution ($E^2\ {\rm d}N/{\rm d}E$) for the gamma-ray emission resulting from neutralino pair annihilation for models A, B and C in Fig.~\ref{fig:dmsources}. In the figure, we do not include the monochromatic line emissions.

\section{DMFIT: Modeling Simulated Dark Matter Sources}\label{sec:dmfitdmonlly}

In this section, we examine fits to the simulated GLAST spectra from dark matter only, prior to considering the full complications of the Galactic center.  This analysis is broadly applicable to any relatively isolated gamma-ray source which may have a dark matter origin (local dwarf galaxies/Galactic subhalos, clusters). First, we consider the specific case of 3EG J1746-2851 and ask whether in one year of data GLAST can distinguish between a dark matter origin (DM model C) and a more typical astrophysical spectrum (Fig.\ref{fig:sga}, Scenario 1), an important question for any unidentified gamma-ray source.  Next, we examine the simulated one year GLAST spectra for DM models A and B, and we use DMFIT to investigate how well we can determine the dark matter particle properties, including both statistical and systematic uncertainties in the particle mass, annihilation cross-section, dominant final state, and final state branching ratios.  Although, we assume a distance and density profile appropriate for the Galactic center, the GLAST exposure time necessary to achieve similar statistics can easily be rescaled for different source and dark matter particle properties.  For example, a highly concentrated, nearby dark matter clump can easily have a luminosity in gamma rays from dark matter annihilation comparable to the Galactic center \cite{Diemand:2006ik,vialactea,baltzwai}. For instance, the brightest clump considered in ref.~\cite{baltzwai} features, for 5 years, the same order of magnitude of photon counts as we consider here. Alternatively, the signal we employ can be thought to refer to a less luminous clump, but with a WIMP model with a larger pair-annihilation rate.

\subsection{Example: 3EG J1746-2851}\label{sec:egret}

Here we address the question of the origin of 3EG J1746-2851 by simulating GLAST observations for both the DM model C and for the broken power-law models (plus exponential cut-off) assumed for the EGRET spectrum of this source (Scenario 1, see Eq.~(\ref{eq:egreta}) and ref.~\cite{MayerHasselwander:1998hg}). As pointed out in ref.~\cite{baltzwai}, the class of astrophysical sources that could be most easily confused with a dark matter annihilation signal are in fact gamma-ray pulsars, whose spectra can be modeled with a power-law plus exponential cut-off (${\rm d}N/{\rm d}E\propto E^{-\Gamma}\exp[-(E/E_c)^\alpha]$) \cite{nelj}. Notice that our spectral model corresponds here, at large energies, to $\Gamma=3.1$, $E_c=30$ GeV and $\alpha=1$. 

\begin{figure}[t]
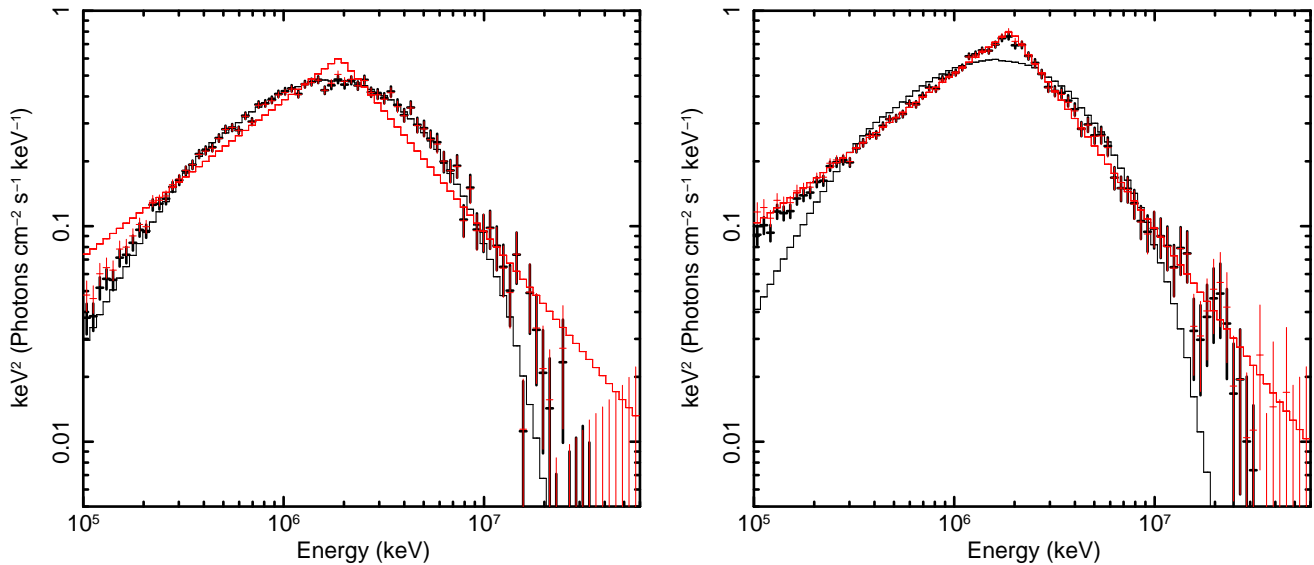

\begin{center}
\mbox{\hspace*{-0.3cm}\includegraphics[height=8.5cm,clip,angle=-90]{FIGURES/dmc_bb_bknp_allE_spec.ps}\quad \includegraphics[height=8.5cm,clip,angle=-90]{FIGURES/ega_bb_bknp_allE_spec.ps}}\\
\caption{Simulated SED for 3EG J1746-2851, assuming the source corresponds to DM model C (left) or to a broken power law plus exponential cutoff (right, see sec.~\ref{sec:saga}), and corresponding XSPEC fits to a DMFIT dark matter model featuring a $b\bar b$ final state (black) and to a broken power law (red).\label{fig:nobckgspectra}}
\end{center}
\end{figure}

The simulated spectra we use here are extracted within a radius of one degree and contain $\sim$ 11,000 and $\sim$ 15,000 photons detected at energies above 1 GeV, respectively.  Fitting the dark matter spectrum ($0.1 - 150$ GeV) with DMFIT in XSPEC to the dominant $b\bar b$ final state, we find a good fit (with a reduced $\chi^2$ of $\chi_{\nu} = 1.17$) and reproduce the particle mass to within 1 GeV ($m_{\chi} = 34.8 \pm 0.5$ GeV).  A broken power-law model for the simulated dark matter spectrum is instead ruled out with $\chi_{\nu} = 6.66$ and a probability $\ll 0.01$.  The one year simulated spectrum for DM model C and the best-fit models are shown in the left panel of Fig.~\ref{fig:nobckgspectra}.  A gamma-ray spectrum of dark matter origin and a broken power-law model are clearly quite different.  Conversely, Fig.~\ref{fig:nobckgspectra}, right, shows the simulated Scenario 1 spectrum for 3EG J1746-2851 and the corresponding best-fit broken power law and DMFIT models.  Again, the correct spectrum is easily identified, with the dark matter model giving $\chi_{\nu} = 7.57$.

Here we have assumed the true dominant final state, $b\bar b$, in our dark matter fits.  However, employing instead a $\tau^+\tau^-$ final state or a $W^+W^-$ final state (with $m_{\chi}$ constrained to be greater than $m_W\approx80$ GeV) gives unacceptable fits to both the DM model C and Scenario 1 spectra, so these final states are effectively ruled out by the data. 

As remarked in ref.~\cite{baltzwai}, a spectrum with a hard power law plus an exponential cutoff, as expected for a gamma-ray pulsar \cite{nelj}, is potentially problematic to distinguish from a dark matter annihilation spectrum, depending on the source brightness. We find that we can tell the two shapes apart, in the case of DM model C (as well as for the fainter DM models A and B) when considering the full energy range $E>0.1$ GeV. Employing a higher energy cutoff, as desirable in the case of high background (see sec.~\ref{sec:angle}), discrimination becomes increasingly problematic, and we recover what found in ref.~\cite{baltzwai}. As also pointed out in ref.~\cite{baltzwai}, other handles beyond a spectral analysis nonetheless exist to discriminate between an astrophysical source such as a gamma-ray pulsar and a dark matter signal. These include for instance source variability (see for instance Ref.~\cite{Atwood:2006kb} and \cite{atwood2}, where  blind searches for a pulsed component in the high energy data were suggested), spatial extent, location and multi-wavelength counterparts \cite{baltzwai,multiw}.

From the example discussed in this section, we see that in one year GLAST will easily determine the spectral shape of 3EG J1746-2851 and, if it is the result of dark matter annihilation, determine the dark matter particle mass with high precision.  Given knowledge of the dark matter density profile, we can also recover the dark matter annihilation cross section to $\sim 10$\% (see sec.~\ref{sec:dmprop2}).  A similar exercise can be used to investigate the possible dark matter origin of any unidentified gamma-ray source.

\subsection{Extracting Particle Properties}\label{sec:dmprop1}

The simulated one year GLAST spectra for DM models A and B have $\sim 2100$ and $\sim 5100$ counts in the energy range $1-150$ GeV and within a radius of 1 degree, respectively. The fact that the dark matter particle mass is significantly higher for these models than for model C and the relatively flat LAT response above 1 GeV \cite{perfpag} mean that we can easily limit the energy range to $> 1$ GeV without losing significant photon statistics (see Fig.~\ref{fig:dmsources}).  Using a higher energy cut is also advantageous when considering the affect of background sources (see sec.~\ref{sec:angle}).  

\begin{figure}[t]
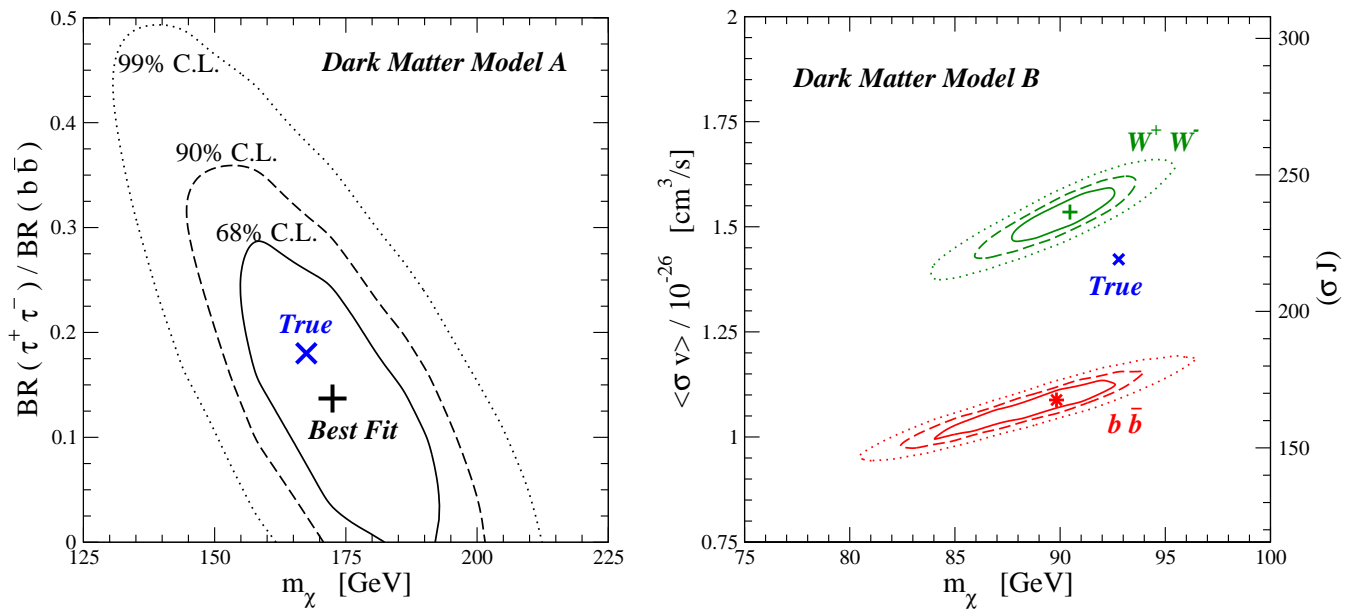

\begin{center}
\mbox{\hspace*{-0.5cm}\includegraphics[height=8.cm,clip]{FIGURES/M_BR.eps}\quad \includegraphics[height=8.cm,clip]{FIGURES/M_SV_B.eps}}\\
\caption{Left: 68\%, 90\% and 99\% confidence level contours for the mass versus the ratio of $\tau^+\tau^-$ and $b\bar b$ final state branching ratios, for a simulated source with an input spectrum given by DM model A. Right: 68\%, 90\% and 99\% confidence level contours on the mass versus normalization $(\sigma\ {\rm J})$ plane, for a simulated source with DM model B spectrum. The left axis shows the pair annihilation cross section, assuming knowledge of $J\simeq 154$ and correcting for the difference in instrumental response for a point source versus an extended source. The green (upper) lines indicate the DMFIT results assuming a $W^+W^-$ final state and the red (lower) lines assuming a $b\bar b$ final state. \label{fig:contours}}
\end{center}
\end{figure}

Considering first DM model A, which is a combination of the $b\bar b$ (84.4\%) and $\tau^+\tau^-$ (15.2\%) final states, we find that the gamma-ray spectrum is well fit by either a single DMFIT model with a $b\bar b$ final state or to a combination of two DMFIT models, one for each final state, keeping the relative normalizations free to vary.  For example, we find reduced $\chi^2$ of 1.03 and 1.02 for these two cases, respectively.  However, for a single $b\bar b$ final state, we systematically overestimate the particle mass by $\sim10$ GeV.  This systematic effect is easy to understand, as the $\tau^+\tau^-$ final state has a significantly harder gamma-ray spectrum \cite{Cesarini:2003nr,Profumo:2005xd}.  For a combination of final states, however, we recover the mass and the branching ratios quite well with $m_{\chi} = 172^{+8}_{-4}$ GeV and $BR(\tau^+\tau^-)/BR(b\bar b) = 0.14 \pm 0.08$.  The degeneracy between final state branching ratio and dark matter particle mass can be seen in the left panel of Fig.~\ref{fig:contours}, showing the confidence level contours on these two parameters.  In this case, the presence of the $\tau^+\tau^-$ final state is not required by the data, but it is clear that if one considers only a single final state the uncertainty on the particle mass is significantly underestimated.  For a brighter source (or a longer integration time) one can actually require a second final state (see sec.~5.2).

To further understand how the unknown final state combination affects the reconstructed particle properties, we consider DM model B, whose dominant final state is $W^+W^-$ (90.4\%) with a small contribution from a $b\bar b$ final state (7.2\%).  Fitting the spectrum to a purely $W^+W^-$ or a purely $b\bar b$ final state, a good fit is found in either case, $\chi_{\nu} = 1.09$ and $1.02$, respectively.  A purely $\tau^+\tau^-$ final state is instead ruled out with $\chi_{\nu} = 5.74$.  As the $W^+W^-$ and $b\bar b$ final states have relatively similar gamma-ray spectral shapes \cite{Cesarini:2003nr,Profumo:2005xd,ds}, they give similar constraints on the particle mass ($m_{\chi} = 90.4^{+1.0}_{-2.0}$ GeV for $W^+W^-$ and $m_{\chi} = 89.8^{+2.0}_{-4.0}$ GeV for $b\bar b$) within $\sim$ 3 GeV and $2 \sigma$ of the true mass.  However, these two final states give significantly different normalizations (respectively, $(\sigma\ {\rm J})$=236 for a purely $W^+W^-$ model and 167 for $b\bar b$), implying, for a given dark matter density profile, two different pair annihilation cross sections (respectively $\langle\sigma {\rm v}\rangle$=1.53 and 1.09$\times 10^{-26}$ cm${^3}/$s).

The DMFIT tool, as implemented in XSPEC, outputs the photon yield in given energy bins for a given annihilation final state, per pair-annihilation event (i.e., the quantity ${\rm d}N_f/{\rm dE}$ of Eq.~(\ref{eq:flux}) integrated over given energy bins). The XSPEC output, therefore, is in units of ${\rm cm}^{-2}\ {\rm s}^{-1}$. To convert this to physically relevant quantities such as the integrated line-of-sight dark matter density squared and the dark matter pair annihilation cross section, we introduce the following adimensional quantity:
\begin{equation}
\left(\sigma \ {\rm J}\right)\equiv J(\psi,\Delta\Omega)\left(\frac{\langle\sigma v\rangle}{10^{26}\ {\rm cm}^3\ {\rm s}^{-1}}\right).
\end{equation}
Calling $\cal O_{\rm XSPEC}$ the XSPEC output normalization, the conversion to the physically relevant quantity $(\sigma\ {\rm J})$ for a WIMP mass $\widetilde m_\chi$ is then simply
\begin{equation}
\left(\sigma \ {\rm J}\right)={\cal O_{\rm XSPEC}}\left(\frac{\widetilde m_\chi}{50\ {\rm GeV}}\right)^2\frac{{\rm cm}^{2}\ {\rm s}^{1}}{3.74\times 10^{-10}}.
\end{equation}
Assuming knowledge of $J(\psi,\Delta\Omega)$, one can then extract the pair annihilation cross section
\begin{equation}
\langle\sigma v\rangle=\frac{\left(\sigma \ {\rm J}\right)}{J(\psi,\Delta\Omega)}\times 10^{26}\ {\rm cm}^3\ {\rm s}^{-1}.
\end{equation}

The right panel of Fig.~\ref{fig:contours} shows the confidence level contours of the particle mass and pair annihilation cross section from DMFIT for DM model B assuming either a $W^+W^-$ or $b\bar b$ final state.  While both models reproduce the true mass relatively well, the true cross section is overestimated by the $W^+W^-$ model and underestimated by the $b\bar b$ model.  Here we have corrected the spectral normalizations for the difference in response between a point source (as assumed by the GLAST tools) and an extended source (see below).  As these two final states have similar gamma-ray spectral shapes, the error on the branching ratio for a combination of the two final states is large, and we find that we can not independently constrain the branching ratio in this case.  The total uncertainty in the annihilation cross section is, therefore, represented by the approximately 30\% offset in the cross sections detemined for a  $W^+W^-$ versus a $b\bar b$ final state, as shown in the right panel of Fig.~\ref{fig:contours}.  If we fix the ratio of these two final states to the true value, the true pair annihilation cross section is reproduced.

In the reconstruction of the cross section from the XSPEC normalization as well as the derivation of source flux and luminosity, another very important systematic  effect needs to be emphasized. In the extraction of instrumental response files appropriate for use with XSPEC, the GLAST tools ({\tt gtrspgen}) assume that the spatial distribution of the source is point-like to account for the reduction in flux versus energy for photons scattered outside of the spectral extraction region based on the instrument point spread function (PSF).  In this section, we use a spectral extraction region of 1 degree radius, which is similar to the GLAST PSF at the lowest energy considered (1 GeV).  However, for a source that is spatially extended on the scale of the extraction region, as is the case for a dark matter source at the Galactic center, significantly more flux is lost to PSF blurring than for a point source.  This effect could be mitigated by taking an extraction region significantly larger than the source extent, but, particularly for the Galactic center, this would come with the price of a significantly increased background (see sec.~5.1).  Dark matter sources other than the Galactic center are expected to have significantly smaller spatial extent also significantly reducing any systematic affect.  Given a known (in the case of simulations) or assumed (in the case of observations) spatial distribution of the source, one can relatively easily estimate the expected reduction in flux for an extended source versus a point source.  Here we have calculated this factor by re-simulating the dark matter source as a point source to find the increase in observed flux in the spectral extraction region and energy range used.  For the DM model B spectrum this factor is roughly 1.6, and the cross sections in the right panel of Fig.~\ref{fig:contours} have been adjusted accordingly.

\section{The Galactic Center Region}\label{sec:gc}
\begin{figure}[t]
\begin{center}
\includegraphics[width=10.7cm,clip]{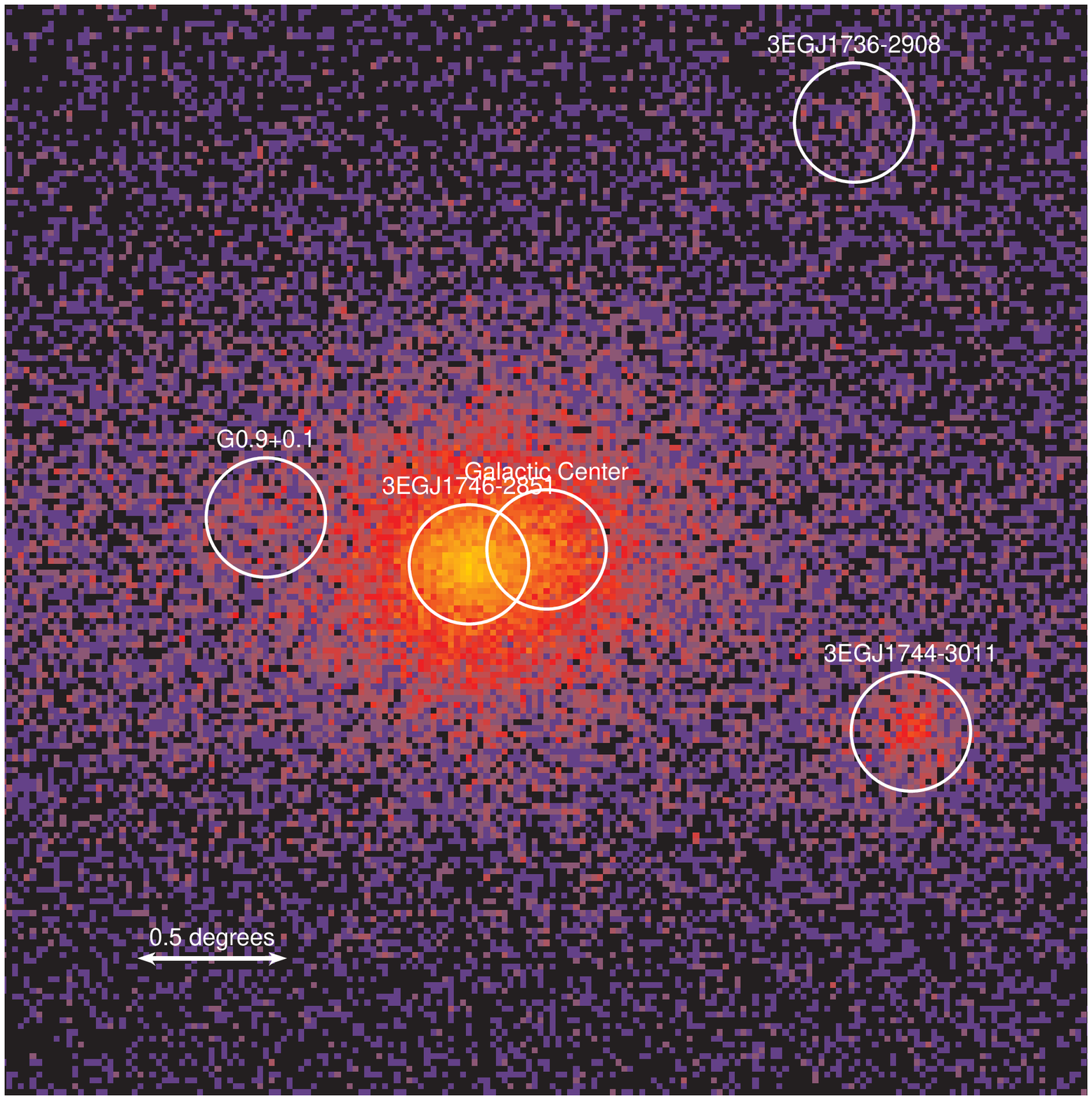}\\[0.5cm]
\includegraphics[width=10.7cm,clip]{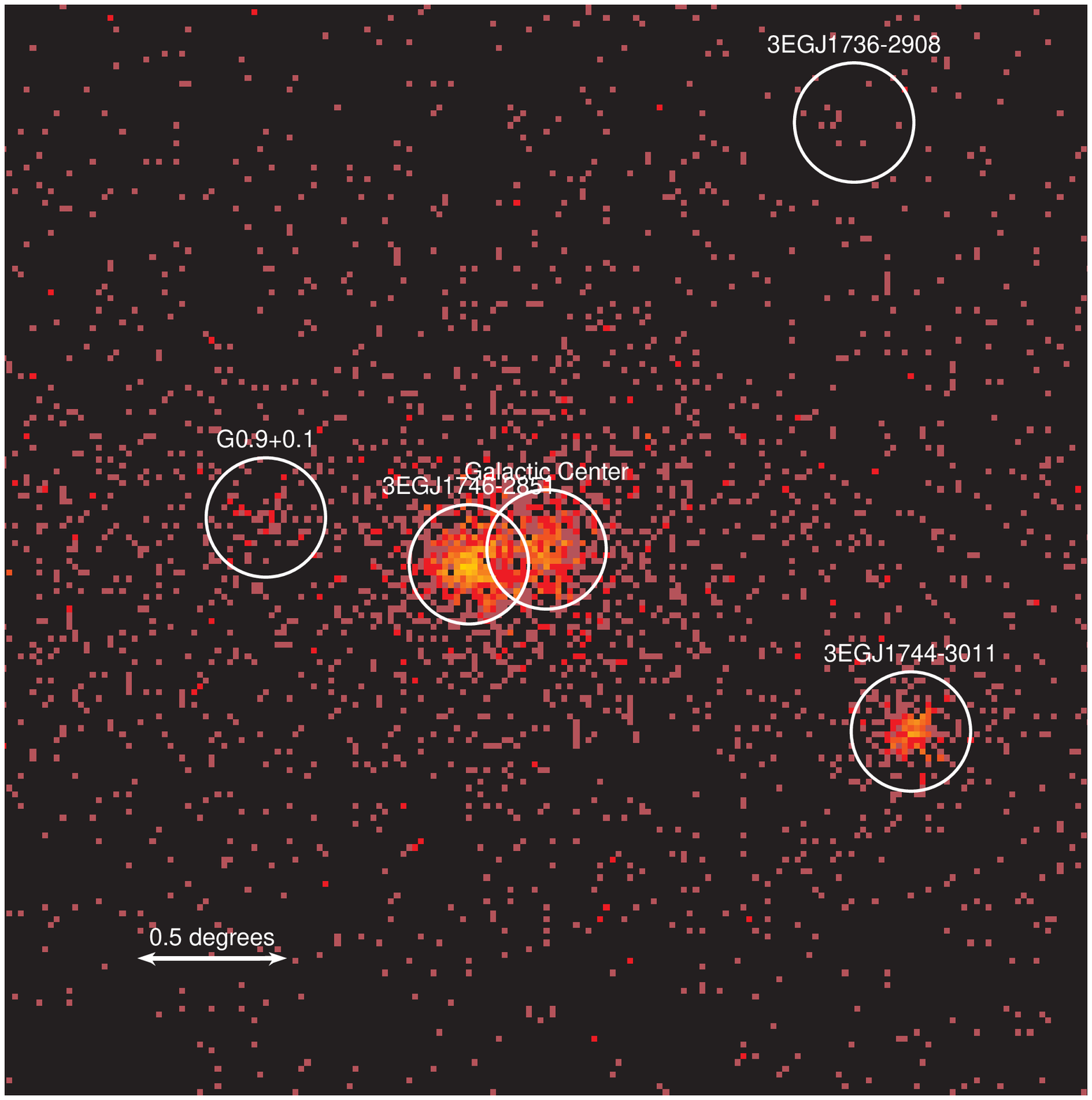}\\
\caption{Simulated gamma-ray images for the inner $3.6^\circ\times3.6^\circ$ of the Galactic center region, for $E_\gamma>1$ GeV (upper panel) and $E_\gamma>5$ GeV (lower panel), assuming background Scenario 1 and DM model B.  The ``Galactic Center'' source includes both the DM source and HESS 1745-290. \label{fig:images}}
\end{center}
\end{figure}

This section is devoted to the analysis of the anticipated gamma-ray sky in the direction of the center of the Galaxy. The full results of our simulations are shown in Fig.~\ref{fig:images}, for $E_\gamma>1$ GeV (upper panel) and $E_\gamma>5$ GeV (lower panel), and assuming background Scenario 1 and DM model B. The figures refer to an angular region of $3.6^\circ\times3.6^\circ$, and all point-sources included in this analysis are highlighted with white circles.

\clearpage

\subsection{The Optimal Energy and Angular Regions}\label{sec:angle}

Prior to embarking on the analysis of the full simulated gamma-ray sky in the Galactic center region and to addressing the question of the potential of GLAST to pinpoint particle dark matter properties (sec.~\ref{sec:dmprop2}), in the present section we investigate the optimal energy and angular cuts from a theoretical standpoint. We show results for all of our dark matter models and background scenarios.

\begin{figure}[t]
\begin{center}
\includegraphics[width=0.95\textwidth,clip]{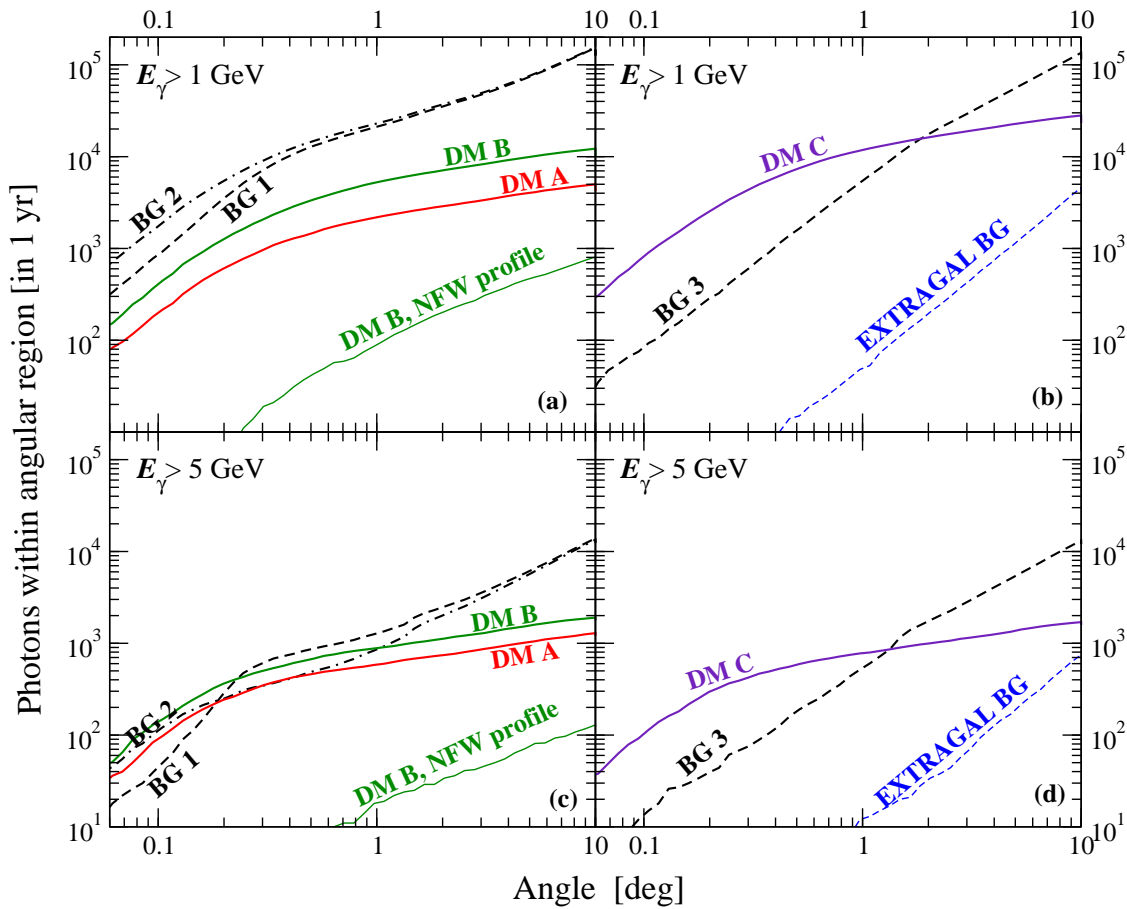}
\caption{Integrated number of photons within a given angular region, for $E_\gamma>1$ GeV (upper panels) and $E_\gamma>5$ GeV (lower panels) for both the total backgrounds and for the DM models. \label{fig:SN_PH}}
\end{center}
\end{figure}
Fig.~\ref{fig:SN_PH} shows the simulation results for the integrated number of photons as recorded by LAT above 1 GeV (upper panels, (a) and (b)) and above 5 GeV (lower panels (c) and (d)), within a given angular region. To produce this figure, we took the simulated event-by-event photon tables and simply summed the number of photons with appropriate energy and position. The panels on the left show the results for background Scenario 1 (two distinct sources for 3EG J1746-2851 and HESS J1745-290, dashed line) and Scenario 2 (single source with curvature radiation inverse Compton emission, dot-dashed line), and for DM model A (red thick solid curve), B (green thick solid curve) and model B, with a NFW profile (lower green thin curve). The right panel refers to background Scenario 3 (HESS J1745-290, dashed line) plus DM model C (indigo solid line).  The backgrounds include all point sources and the Galactic diffuse emission. In the right panels, we also show the integrated photon counts for the extra-galactic gamma-ray background (dashed blue line), which as anticipated only contributes non-negligibly at large angles, and still only at the few percent level.

As expected, at large angles the background is in all cases dominated by the diffuse emission. The estimated number of photons at large angles is very close for all background models. The inclusion of the source associated to 3EG J1744-3011 [HESS J1745-303] manifests itself in the bump visible in the $E>5$ GeV panels (c) and (d) between $1^\circ$ and $2^\circ$. As clear also from Fig.~\ref{fig:images}, G 0.9+0.1 and 3EG J1736-2908 play a less significant role. The different assumptions on the gamma-ray sources associated to 3EG J1746-2851 and HESS J1745-290 are evident in angular regions below $0.5^\circ$. In particular, background Scenario 1, where 3EG J1746-2851 is $\sim0.2^\circ$ off-set from the Galactic center, rapidly increases and overtakes Scenario 2 in the total number of photons above 5 GeV right around $0.2^\circ$, as expected. Background Scenario 2, on the other hand, has a larger number of photons with energies above 1 GeV at all angular scales.

Since we assume the same dark matter profile, modulo statistical fluctuations, the dark matter annihilation signals from models A, B and C are obviously proportional to each other. The signal is dominated by the innermost region, and the increase at large angular scales (due to both the instrumental PSF and to the actual annihilation events at large radii) is significantly less steep than the background. This does not hold for the case of a NFW profile, where the signal (as expected from the NFW profile functional form, $\rho_{\rm DM}\propto r^{-1}$) increases linearly with angle. The innermost region is, therefore, less crucial for a NFW, as we will also detail and quantify below.

Comparing the simulated photon counts above 1 and 5 GeV, it is clear that the ratio of the signal to the square root of the background event counts ($S/\sqrt{B}$) is more favorable for the $>$1 GeV case than for the $>$5 GeV case for DM model C and background Scenario 3. Quantitatively, we find that $S/\sqrt{B}$ is larger by a factor $\sim4$ over almost the entire set of angular regions we consider. The situation is less clear-cut for background Scenarios 1 and 2 with our reference dark matter profile. In the case of background Scenario 1, the $S/\sqrt{B}$ is larger above 5 GeV in small angular regions (namely below $0.2^\circ$, where 3EG J1746-2851 kicks in), while it is larger for the 1 GeV cut for larger angular regions. Similarly, for background Scenario 2 we find a slightly larger $S/\sqrt{B}$ for the 5 GeV cut out to $\sim 0.75^\circ$. In a $5^\circ$ region the $S/\sqrt{B}$ for the 1 GeV cut is instead larger by a factor $\sim2$ compared to the 5 GeV cut. For the case of the NFW profile, $S/\sqrt{B}$ for the 1 and 5 GeV cuts is essentially the same in angular regions smaller than $\sim1^\circ$, while it is significantly larger for the 1 GeV cut in larger angular regions.

The optimal energy range and angular region, however, does not only depend on $S/\sqrt{B}$: considering a lower energy cut significantly increases the statistics in the number of signal events and carries additional important spectral information. For the case of a NFW profile, the 5 GeV cut simply limits the statistics of the number of signal events unacceptably (compare panels (a) and (c)). Considering small angular regions and a 5 GeV cut also appears to limit the number of photon events in such a way as to hinder the optimal reconstruction of dark matter particle properties: for instance, looking at the DM model A and B lines in panels (a) and (c), $~100$ signal photons for a 5 GeV cut within $0.1^\circ$ carry significantly less information on particle dark matter properties than $3-6\times 10^3$ photons above 1 GeV and within $1^\circ$.

\begin{figure}[t]
\begin{center}
\includegraphics[width=0.65\textwidth,clip]{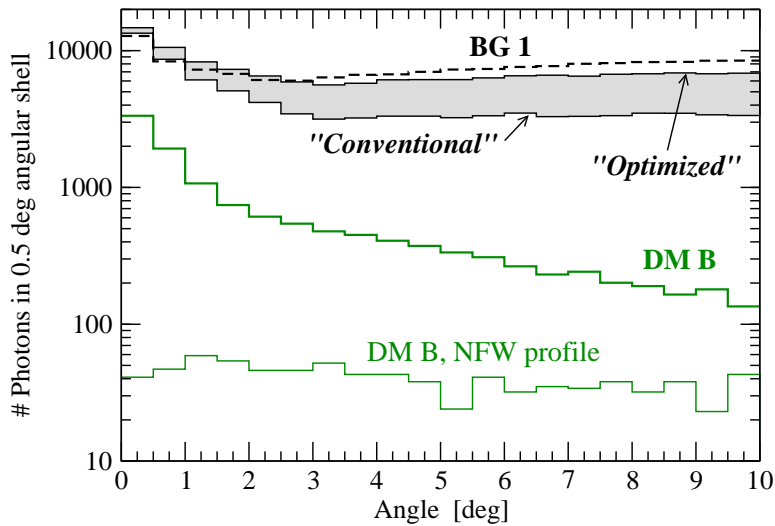}
\caption{Integrated number of photons, with $E_\gamma>1$ GeV, in spherical shells of $0.5^\circ$, at given angular distances. The gray band indicates the range between the ``conventional'' and ``optimized'' backgrounds described in ref.~\cite{Baltz:2008wd}. \label{fig:SHELL}}
\end{center}
\end{figure}
In view of these considerations, we decided to adopt a 1 GeV cut. Fig.~\ref{fig:SHELL} illustrates the number of photon counts inside shells of $0.5^\circ$, between 0 and $10^\circ$, for background Scenario 1 (dashed black line) and for DM model B with our reference dark matter profile (upper solid green line) and with a NFW profile (lower green line). For comparison, we also include the ``optimized'' (upper boundary of the gray band) and ``conventional'' (lower) background models of ref.~\cite{Baltz:2008wd}. Notice that the number of background photons within spherical shells appears to be at a minimum between 2 and 3 degrees (3 and 4 for the background setups of ref.~\cite{Baltz:2008wd}), and it keeps increasing for shells at larger angles. With our standard dark matter profile, we find that considering circular annular shells around the Galactic center region does not help in increasing the signal-to-noise; for a NFW profile, we find that an annulus between 2 and 3 degrees can actually have a larger $S/\sqrt{B}$, although with only a few hundreds of signal photon counts.

We also tried alternative angular cuts, such as for instance a cut in Galactic latitude, that would allow us to remove part of the Galactic diffuse background. We find that with this cut topology the signal-to-noise does not improve neither for a NFW nor for a steeper dark matter profile.

\begin{figure}[t]
\begin{center}
\includegraphics[width=1.0\textwidth,clip]{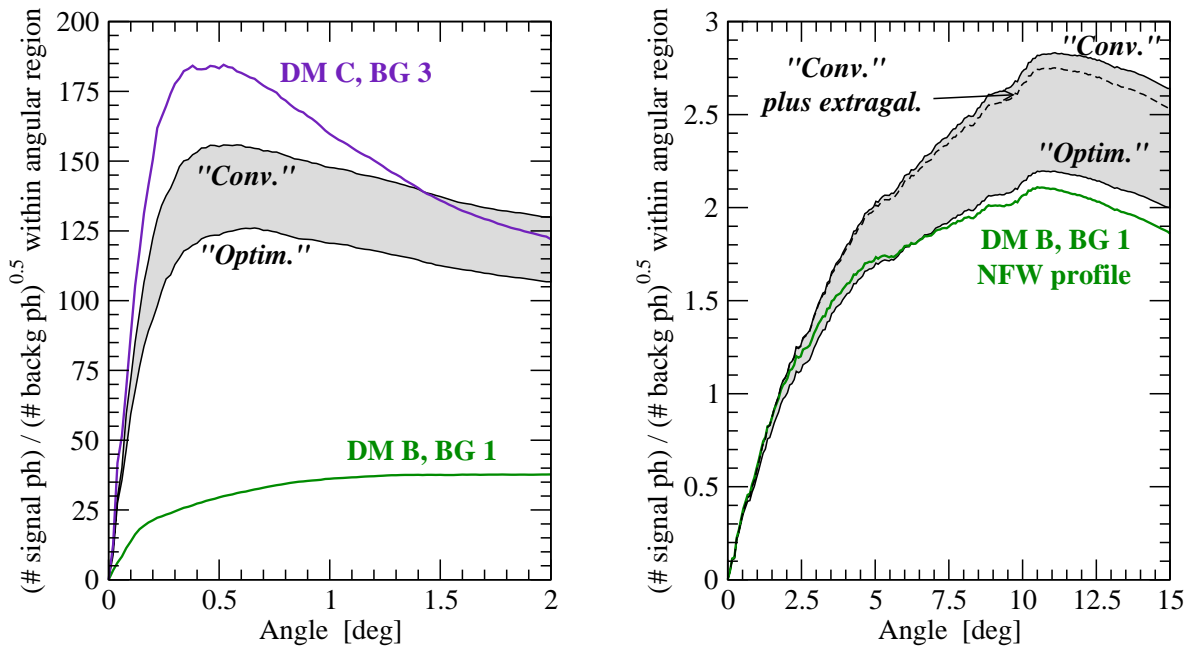}
\caption{Ratio of the integrated number of $E_\gamma>1$ GeV photons within given angular regions, over the square root of the number of background photons in the same region (i.e. an estimate of the ``significance'', or signal-to-noise ratio) for DM model C with background Scenario 3 and DM model B with background Scenario 1 (left), and DM model B, background Scenario 1 and a NFW profile (right). The gray bands indicate the ranges between the ``conventional'' and ``optimized'' backgrounds described in ref.~\cite{Baltz:2008wd}. The dashed line in the right panel includes the extra-galactic gamma-ray background. \label{fig:SN}}
\end{center}
\end{figure}
Having assessed that leaving out the innermost angular regions is not an optimal strategy, we show in Fig.~\ref{fig:SN} the total $S/\sqrt{B}$ within given angular regions, for a few choices of background and DM models and with an energy cut at 1 GeV. In particular, the left panel shows $S/\sqrt{B}$ for DM model C with background Scenario 3 (upper indigo line) and for DM model B with background Scenario 1 (lower green line). In the case where 3EG J1746-2851 is actually dominantly of dark matter origin (DM model C, background Scenario 3), we find that the $S/\sqrt{B}$ peaks at an angular region within $0.5^\circ$ from the Galactic center. This statement doesn't depend upon the diffuse background model (in this case the dominant source of background). The gray band indicates the signal-to-noise for the ``conventional'' (upper curve) and for the ``optimized'' background setups outlined in ref.~\cite{Baltz:2008wd}. Even if at small angular region $S/\sqrt{B}$ decreases, it still peaks in an angular region of around $0.5^\circ$.

In the case where instead 3EG J1746-2851 is part of the background, the total signal-to-noise increases up to $\sim1^\circ$, and then flattens out. At even larger angular regions, $S/\sqrt{B}$ starts to decrease, particularly significantly at angles larger than 3-4 degrees. Since including the region between 1 and 2 degrees only marginally improves the $S/\sqrt{B}$ at the cost of adding $\sim 10^4$ extra background photons (see Fig.~\ref{fig:SHELL}) and less than $10^3$ signal photons, we consider as the optimal angular region a circle of 1 degree around the Galactic center for background Scenarios 1 and 2. For the case of a NFW profile, instead, (right panel), we find that the largest $S/\sqrt{B}$ is achieved at around 10 degrees. Notice however that this does not include additional gamma-ray sources between 4 and 10 degrees that might further degrade the $S/\sqrt{B}$ at large angles. Again, the gray band highlights that this assessment is independent of the detailed model for the diffuse background. The dashed line indicates how $S/\sqrt{B}$ is affected by including the extra-galactic background: as already pointed out above, including it does not affect in any significant way our results, nor the optimal angular region.

In summary, we find that the optimal energy cut is likely around 1 GeV, and that the optimal angular region includes the innermost part of the Galactic center and depends on the assumed dark matter profile. For a steeply rising profile, the optimal angular range is between $0.5$ and $2$ degrees, while for a NFW profile it extends to significantly larger angles, possibly up to around 10 degrees.

In view of the above considerations, in the following analysis we will use an angular region of 0.5 degrees for DM model C and background Scenario 3, and of 1 degree for DM models A and B and background Scenarios 1 or 2.  For background Scenarios 1 and 2 we will use only photons above 1 GeV, while for background Scenario 3 and DM model C we extend the spectral range down to photon energies of 100 MeV.

\subsection{Dark Matter Particle Properties from the Galactic Center}\label{sec:dmprop2}

We now investigate whether we can significantly detect the spectral signature of dark matter annihilation and constrain dark matter particle properties in one year of GLAST data from the Galactic center in the three background Scenarios detailed in sec.~\ref{sec:saga}.  Here all sources and the Galactic diffuse emission are included in the GLAST simulations.
\begin{figure}[t]
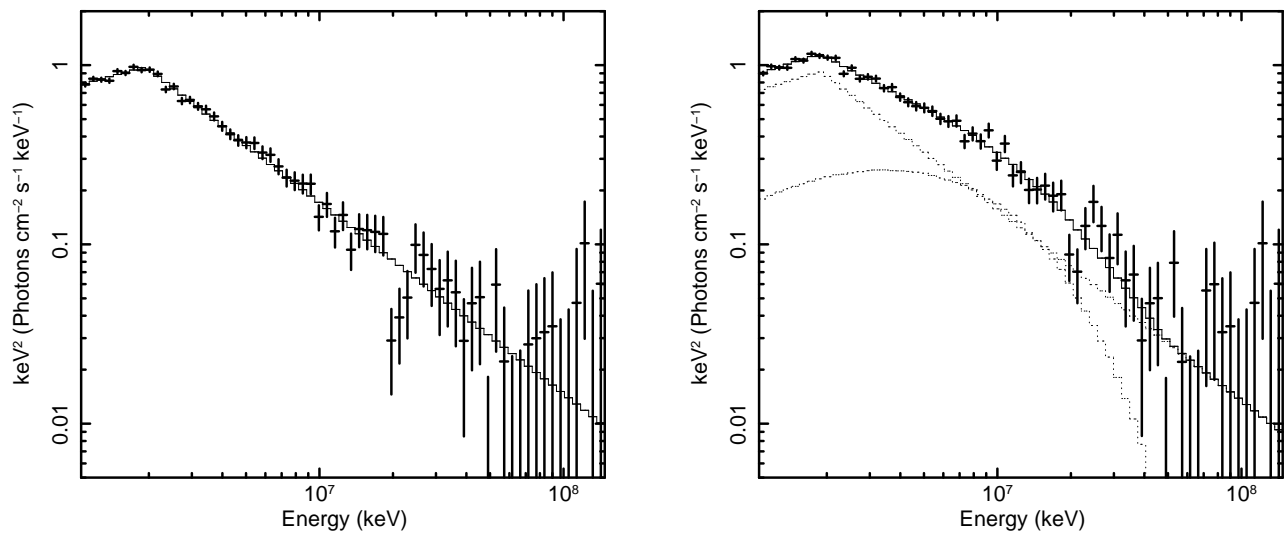

\begin{center}
\mbox{\includegraphics[width=7.cm,clip,angle=-90]{FIGURES/bkgd1_bknp_spec.ps}\qquad\ \includegraphics[width=7.cm,clip,angle=-90]{FIGURES/dmbbkgd1_bknpdmfit_spec.ps}}\\
\caption{Full simulation, in an angular region of $1^\circ$, for background Scenario 1 only (left) and for background Scenario 1 plus DM model B (right).  The background spectrum is modeled as a broken power law.\label{fig:spec1}}
\end{center}
\end{figure}
\begin{figure}[t]
\begin{center}
\mbox{\includegraphics[width=7.cm,clip,angle=-90]{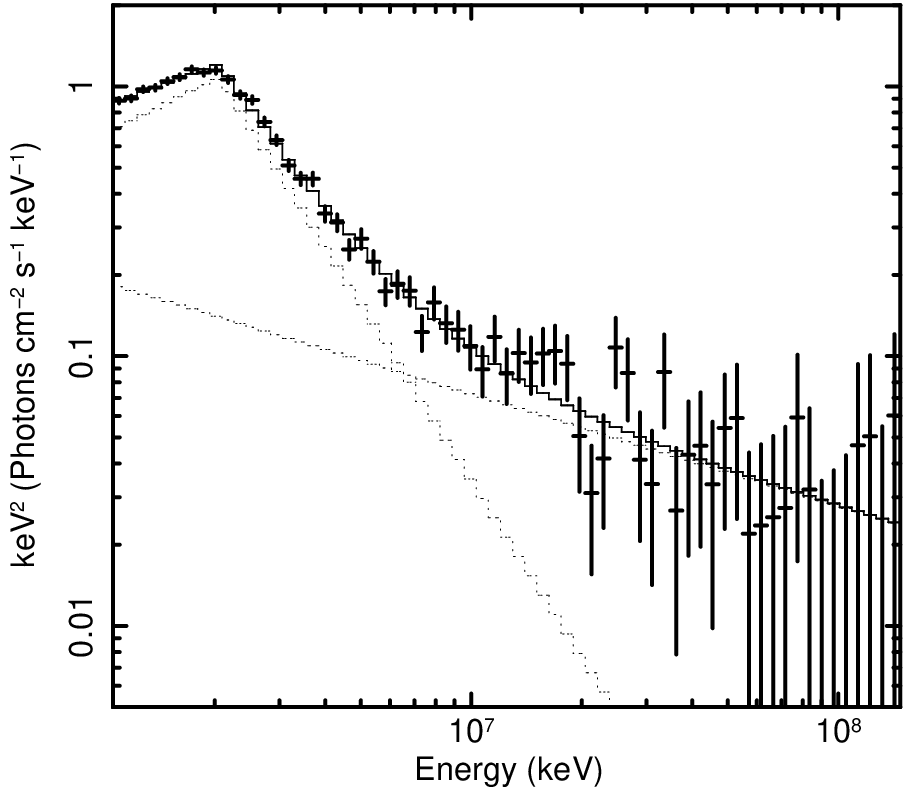}\qquad\ \includegraphics[width=7.cm,clip,angle=-90]{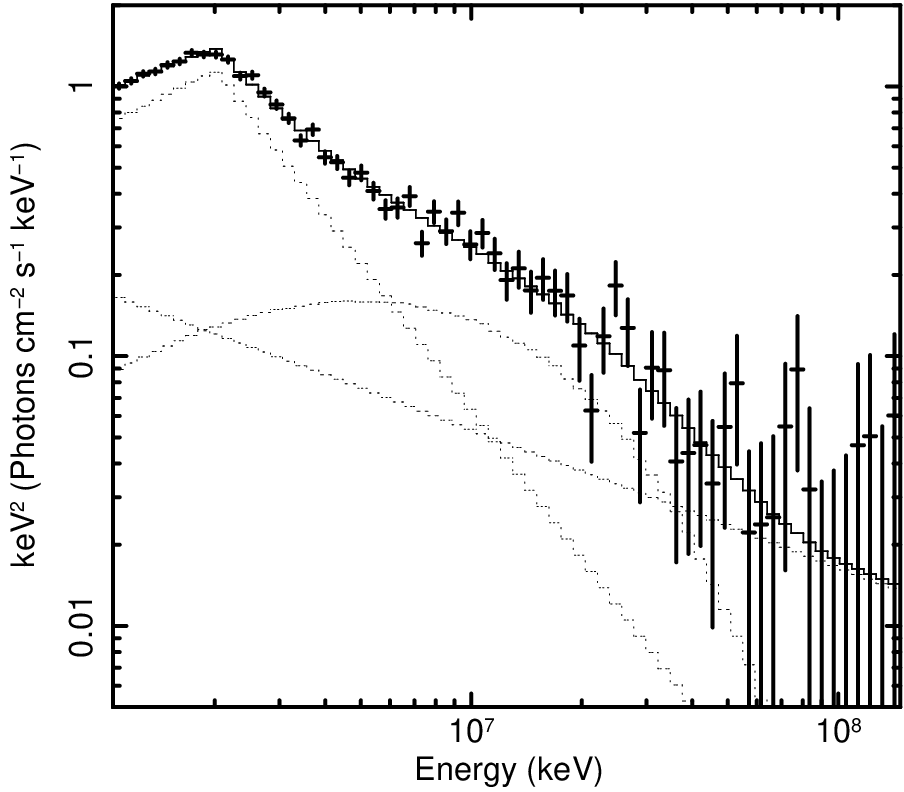}}\\
\caption{Full simulation, in an angular region of $1^\circ$, for background Scenario 2 only (left) and for background Scenario 2 plus DM model B (right).  The background spectrum is modeled as a broken power law plus a single power law.\label{fig:spec2}}
\end{center}
\end{figure}
\begin{table}
\caption{Summary of fits to the dark matter annihilation spectrum in the 
Galactic center region for background Scenarios 1 and 2. The first two 
columns indicate the dark matter and background Scenarios, and the 
remaining columns list the reduced $\chi^2$ of the fits with the dark 
matter particle mass in brackets when a dark matter component is present. 
Column 3 indicates the fit to the spectrum of the background only.  This fit is 
to a broken power-law model for Scenario 1 and to a broken power law 
plus an additional power law for Scenario 2 (see \ref{sec:dmprop2}).  The remaining 
columns show the fits to the background plus dark matter spectrum assuming 
the background model only (Col.~4) and a background plus DMFIT model 
(Cols. 5 and 6).  In column 6 the background model parameters are fixed at 
their best fit values from the background only fit, while in the other 
columns all model parameters are allowed to vary.  Models with a 
probability less than 1\% are shown in bold.\\[0.1cm] \label{tab:GCF}}
\begin{indented}
\item[]\hspace*{-2.75cm}\begin{tabular}{|ccc|c|ccc|}
\hline
  Bkgd & DM & DM & Bkgd Only & & Bkgd + DM Spectrum & \\
  Model & Model & Mass& Spectrum & bkgd spec. & bkgd spec. + DMFIT & bkgd spec. 
(fixed) + DMFIT\\
\hline
{\quad }&  & & & & & \\[-0.2cm]
1 & A & {\color{red}167.5} &1.22 &1.20  &1.10\quad [{\color{red}$144^{+30}_{-40}$}] &1.09\quad [{\color{red}$173^{+16}_{-8}$}] \\[0.15cm]
1 & B & {\color{red}92.8} &1.22 &\textbf{2.86}  &1.36\quad [{\color{red}$81.8^{+3.4}_{-8.7}$}] &1.39 \quad
[{\color{red}$90.6^{+2.0}_{-2.0}$}] \\[0.15cm]
2 & A & {\color{red}167.5} &0.99 &1.24 &0.946\quad [{\color{red}$274^{+56}_{-58}$}] &0.891 \quad
[{\color{red}$208^{+16}_{-16}$}] \\[0.15cm]
2 & B & {\color{red}92.8} &0.99 &\textbf{1.56} &1.29\quad [{\color{red}$109^{+8}_{-16}$}] &1.21 \quad
[{\color{red}$92.4^{+2.0}_{-2.0}$}] \\[0.15cm]
\hline
\end{tabular}
\end{indented}

\end{table}

\begin{itemize}
\item {\bf Scenarios 1 and 2}.  For background Scenario 1, where the EGRET and HESS sources near the Galactic center are different, the EGRET source dominates and the background spectrum can be fairly well modeled by a simple broken power law as shown in Fig.~\ref{fig:spec1} (left) and Tab.~\ref{tab:GCF}.  The reduced $\chi^2$ of this fit is 1.22 due to a small contribution from the Galactic diffuse emission but has a probability greater than 10\%.  We found a significantly better fit to this background spectrum if we either limited the energy range to $> 5$ GeV, which unfortunately leads to a significant reduction in statistics for the dark matter analysis, or if we subtracted off the diffuse background spectrum in XSPEC as might be done observationally by taking a diffuse spectrum from another part of the sky.  Adding an additional power law to model the diffuse component did not improve the fit, and as we prefer to make as few assumptions about the background spectrum as possible, we decided to model the background for Scenario 1 as a single broken power law.  Instead, for background Scenario 2, where the EGRET and HESS sources are the same, an additional power law (i.e.~a broken power law plus power-law model) is necessary to model the gamma-ray spectrum at high energies as shown in Fig.~\ref{fig:spec2} (left).  This model leads to an excellent fit to background Scenario 2 with $\chi_{\nu} = 0.99$ (Tab.~\ref{tab:GCF}).

We then add a dark matter source with either model A or B and investigate the necessity of a dark matter component in the spectral fit and our ability to constrain its parameters.  We first make minimal assumptions about the background spectrum, assuming only the rough spectral shape (a broken power law for background Scenario 1 and a broken power law plus power law for background Scenario 2) and leaving all of the spectral parameters free to vary in the fits.  In practice, some independent information on the spectrum of background sources will be known from other wavelengths or can be derived if the sources are spatially separated (e.g.~the Galactic diffuse and the EGRET source in Scenario 1).  Therefore, we also investigate our ability to fit the dark matter source if we fix the background model at its best fit.  The reduced $\chi^2$ and constraints on the dark matter mass for all fits are listed in Tab.~\ref{tab:GCF}.

For both background Scenarios and DM model B, a background only spectrum with no dark matter component is ruled out at better than 99\% confidence level even without assuming the background spectral parameters.  The addition of a DMFIT model with the dominant $W^+W^-$ final state significantly improves the fit, showing that we can detect the presence of a dark matter source even if we do not know the precise background (Fig.~\ref{fig:spec1} and Fig.~\ref{fig:spec2}, right).  However, with the inclusion of the background, the best-fit particle mass is shifted systematically to lower values for background Scenario 1 and to higher values for background Scenario 2 by $10-20$ GeV, but within a few sigma.  If we instead fix the background model at its best fit, this systematic shift disappears, and the statistical errors on the mass are significantly reduced.  Some knowledge of the background spectral shape clearly improves the accuracy in the determination of dark matter particle properties.

For a dark matter source with DM model A the flux is lower, and a dark matter component is not clearly necessary in the spectral fits.  Adding a dark matter component with the dominant $b\bar b$ final state does, however, improve the fits, and in the case background Scenario 2 the low energy slope of the broken power law is significantly different than its true value if no dark matter component is included.  As before, we find that the best-fit particle mass is too low for background Scenario 1 and too high for background Scenario 2, although the large errors on the mass incorporate these shifts.  With the background models fixed at their best fits, we again recover the particle mass with much better accuracy.  Here the best-fit masses are offset high for both backgrounds, at least partially due to the fact that we have not included the subdominant $\tau^+\tau^-$ final state.

\item {\bf Scenario 3}.  Here we reconsider the analysis in sec.~\ref{sec:egret} for the case where the EGRET source 3EG J1746-2851 is due to dark matter annihilation with DM model C, but now we include the backgrounds.  As we take a relatively small spectral extraction region (radius 0.5 degrees), the only significant background sources are the Galactic diffuse and at higher energies the HESS source, HESS J1745-290.  Due to the lower dark matter particle mass for model C as well as the lower background, we find it advantageous to extend the spectrum down to lower energies, and here we use a spectral energy range of $0.1-150$ GeV.  Because of the larger GLAST PSF at lower energies, the surrounding point sources outside of the spectral extraction region will contribute some flux at the lower end of the spectral energy range.

\begin{figure}[t]
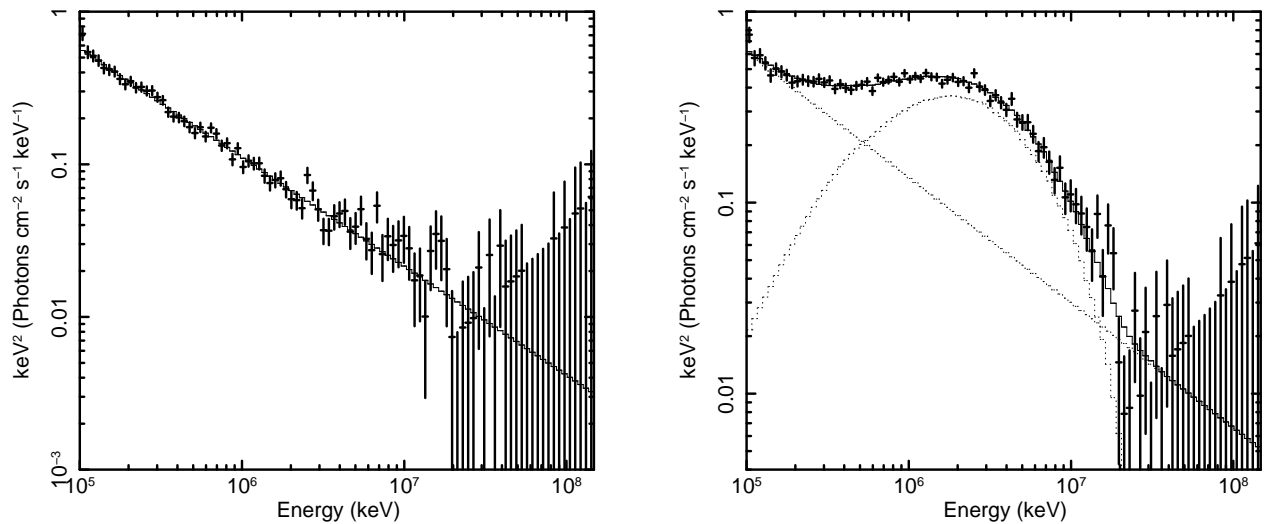

\begin{center}
\mbox{\includegraphics[width=7.cm,clip,angle=-90]{FIGURES/bkgd3_power_spec.ps}\qquad\ \includegraphics[width=7.cm,clip,angle=-90]{FIGURES/dmcbkgd3_pdmfit_allE_spec.ps}}\\
\caption{Full simulation, in an angular region of $0.5^\circ$, for background Scenario 3 only (left) and with DM model C (right), with a background model consisting of a single power law.\label{fig:spec3}}
\end{center}
\end{figure}
As shown in Fig.~\ref{fig:spec3} (left), the background only spectrum is well fit by a simple broken power law with a spectral index of 2.7 ($\chi_{\nu} = 0.88$).  Considering the full background plus DM model C spectrum, we again ask the question whether we can distinguish between a source with a dark matter-like spectrum modeled with DMFIT versus a broken power-law spectrum.  Again, we find that a broken power-law source spectrum is ruled out at better than 99\% confidence level ($\chi_{\nu} = 1.46$ for 90 DOF) even if the shape of the background power law is assumed to be completely unknown, in which case the background power-law slope is much different than its best fit ($\Gamma = 4.9$).  A background power law plus DMFIT model with the dominant $b\bar b$ final state, on the other hand, gives and excellent fit, shown in Fig.~\ref{fig:spec3} (right), with $\chi_{\nu} = 0.84$ and $m_{\chi} = 36.4^{+0.6}_{-0.8}$ GeV.  In this case, the background power-law slope is also reproduced within the errors.  With the background model fixed at its best fit, we similarly find $\chi_{\nu} = 0.95$ and $m_{\chi} = 35.3^{+0.5}_{-0.5}$ GeV.  

\begin{figure}[t]
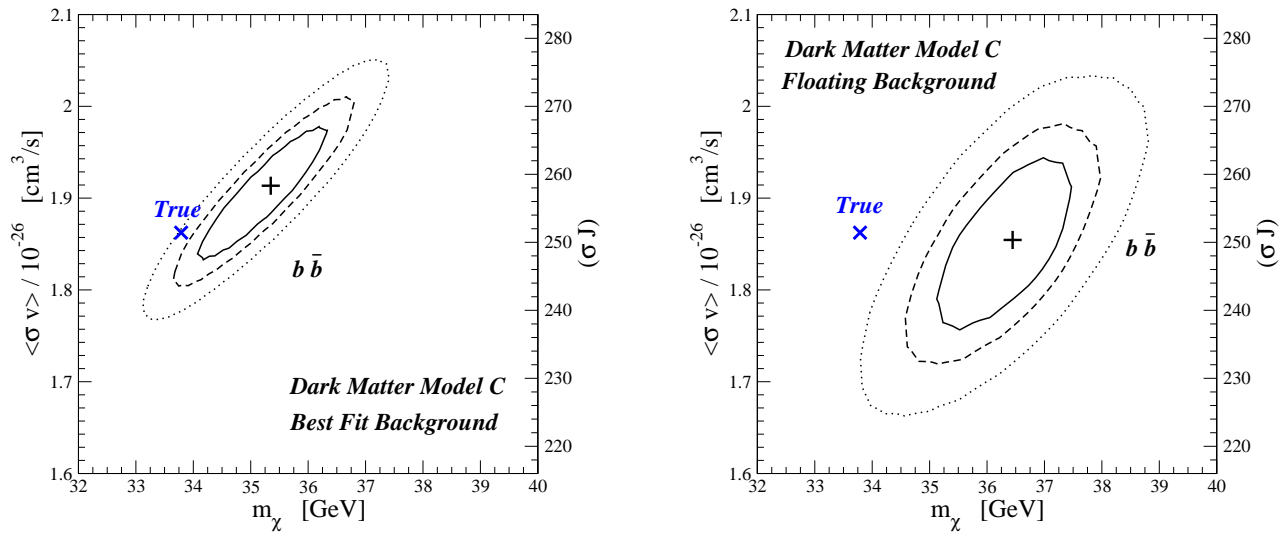

\begin{center}
\mbox{\includegraphics[height=7.cm,clip]{FIGURES/M_SV_C_fix.eps}\qquad \ \ \includegraphics[height=7.cm,clip]{FIGURES/M_SV_C_float.eps}}\\
\caption{68\%, 90\% and 99\% confidence level contours on the mass versus normalization $(\sigma\ {\rm J})$ plane, for DM C with background Scenario 3, in an angular region of $0.5^\circ$. The left axis shows the pair annihilation cross section, assuming knowledge of $J\simeq 67$. The cross indicates the best fit values, the blue ``X'' the true input values. In the left panel, the background is frozen to its best fit power-law model, in the right panel the background model parametrization is a variable in the fit. \label{fig:M_SV_C}}
\end{center}
\end{figure}
\begin{figure}[t]
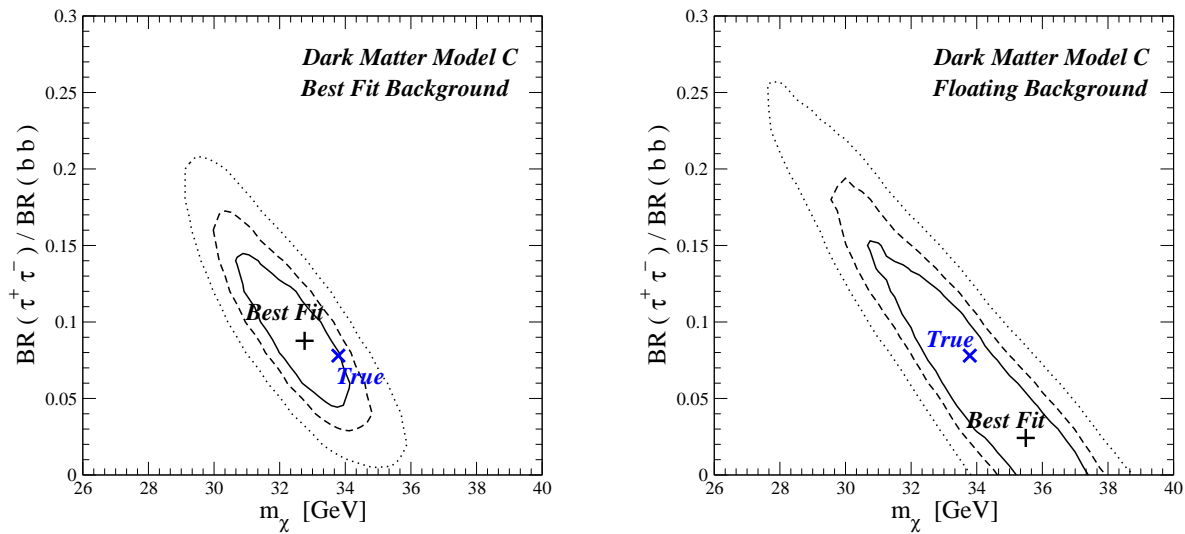

\begin{center}
\mbox{\includegraphics[height=7.cm,clip]{FIGURES/M_BR_C_fix.eps}\qquad \ \ \includegraphics[height=7.cm,clip]{FIGURES/M_BR_C_float.eps}}\\
\caption{68\%, 90\% and 99\% confidence level contours for the mass versus the ratio of $\tau^+\tau^-$ and $b\bar b$ final states for DM C with background Scenario 3, in an angular region of $0.5^\circ$. The cross indicates the best fit values, the blue ``X'' the true input values. In the left panel, the background is frozen to its best fit power-law model, in the right panel the background model parametrization is a variable in the fit. \label{fig:M_BR_C}}
\end{center}
\end{figure}
Fig.~\ref{fig:M_SV_C} shows a comparison of the confidence level contours for the dark matter particle mass and annihilation cross section for the case where we fix the background power law at the best fit for the background only versus the case where we fit for the power-law parameters.  We have again corrected the spectral normalization, and, therefore, pair annihilation cross section, for the flux lost for an extended source versus a point source.  For this model and spectral extraction region, the correction factor was 1.63.  For an unknown background model, the confidence level regions, of course, widen, and we find a small shift within about $2 \sigma$ of the best fit to higher masses and lower cross sections.  In both cases, the true cross section is reproduced well and the mass is offset high compared to the true mass by a couple of GeV, mostly because we have not included the contribution of the subdominant $\tau^+\tau^-$ final state.  

The constraints on the final state branching ratios versus mass when including a $\tau^+\tau^-$ final state are shown in Fig.~\ref{fig:M_BR_C} for both a fixed and fit background model.  In both cases, the true particle mass and branching ratio is reproduced within the errors.  If the background model is known independently, as could easily be the case for the Galactic diffuse, we find that a $\tau^+\tau^-$ final state is required at more than 99\% confidence level (Fig.~\ref{fig:M_BR_C} (left)).

\end{itemize}

In summary, for a bright dark matter source like DM model C, one year of GLAST data will make it possible to constrain the dark matter particle mass to $\sim$ 10\% accuracy (systematic and statistical) even if the background model and final state branching ratios are unknown.  For fainter dark matter sources like models A and B with bright background sources nearby, the particle mass can be determined to similar accuracy if the background spectrum is independently known.  If the background is completely unknown, we derive particle masses with roughly 60\% and 20\% and accuracy for DM models A and B, respectively.

\begin{figure}[t]
\begin{center}
\includegraphics[height=6.2cm,clip]{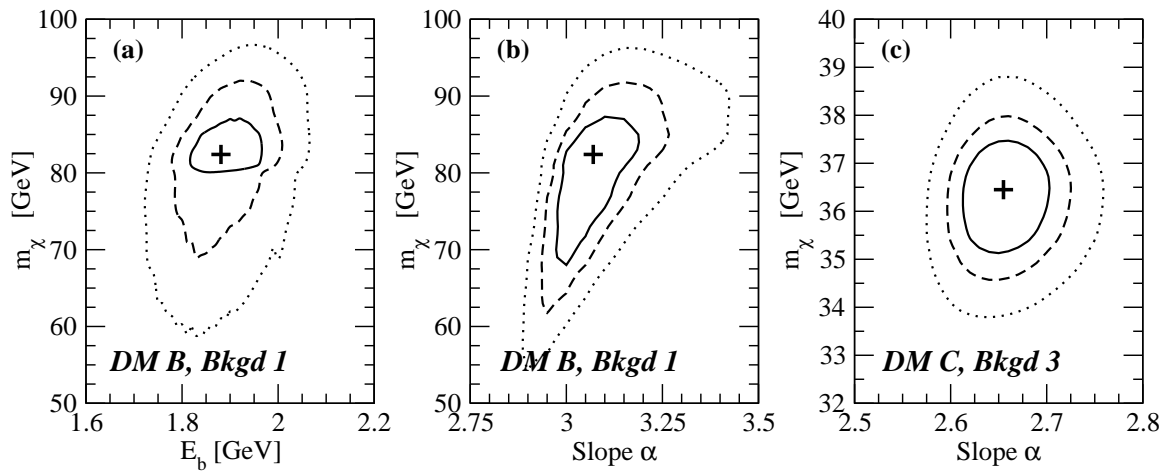}\\
\caption{68\%, 90\% and 99\% confidence level contours for the dark matter mass (on the $y$ axis) versus background parameters. Panels (a) and (b) refer to background Scenario 1 and DM model B, while (c) to background Scenario 3 and DM model C. Panel (a) correlates the mass with the location of the break in the power law adopted for the fit to the background; (b) and (c) correlate the mass with the high-energy slope of the background power-law model. \label{fig:ASTRO}}
\end{center}
\end{figure}
As a final test, we looked for correlations between the parameters of the background spectral model and the DMFIT particle mass.  We limit this study to DM models B and C.  For example, Fig.~\ref{fig:ASTRO} shows the confidence level contours on the particle mass versus high energy slope and break energy for background Scenario 1 and DM model B as well as for the DM model C mass versus the background Scenario 3 power-law slope.  We find no strong correlations between dark matter mass and the background parameters, showing that these are constrained fairly independently.  The most significant trend is a weak correlation, particularly at lower masses, between the high energy slope of background Scenario 1 and the DM B particle mass. We also simulated other diffuse background models, including the ``optimized'' GALPROP and the ``conventional'' diffuse models discussed in ref.~\cite{Baltz:2008wd}. For instance, the optimized model gives approximately an 80\% higher diffuse flux above 1 GeV within 1 degree of the Galactic center.  We find that  our results are quite insensitive to the diffuse model. With the optimized diffuse model, we find for DM model C and background Scenario 3 very similar uncertainties in the determination of the neutralino mass. The best fit mass is biased towards lower values, although by less than 4 GeV. This is well within the uncertainties induced by including more than one final state annihilation mode for our reference diffuse background, see e.g. the right panel of fig.~\ref{fig:M_BR_C}. The impact on the particle dark matter parameter determination for background Scenarios 1 and 2 is even smaller, since there the dominant background stems from the point sources in the Sag A${^*}$ region.

\section{Summary and Conclusions}\label{sec:concl}
With the detailed analysis of a specific example, the Galactic center region, this study reaffirms and reinforces the point that GLAST has the potential to play a pioneering role in the race towards the discovery of the fundamental nature of dark matter. While eagerly awaiting real data to go beyond simulations, the present theoretical study highlighted that:
\begin{itemize}
\item the nature of the EGRET source in the Galactic center, and its possible association to dark matter annihilation, will be conclusively probed by GLAST with less than one year of data;
\item the determination of the dark matter particle properties such as the mass and the dominant and sub-dominant annihilation modes is possible to a varying degree of accuracy, depending on the brightness of the source and on the knowledge of the background spectrum;
\item for realistic particle dark matter models, GLAST has the potential to (i) firmly establish the presence of a dark matter source in the Galactic center, even when it is not the brightest source in the region, (ii) estimate the dark matter particle mass to better than 10\%, and (iii) to disentangle the occurrence of more than one annihilation mode to 99\% confidence level;
\item if the EGRET source at the Galactic center is associated to dark matter annihilation, we showed that the optimal energy range includes all photons above 0.1 GeV within an angular region of $0.5^\circ$; otherwise, the optimal energy range for dark matter searches is above 1 GeV, and the optimal angular region goes from $1^\circ$ for steep dark matter profiles, to $\sim10^\circ$ for shallower profiles;
\item the finite spatial extent of the dark matter source at the Galactic center amounts to a systematic effect in the estimate of the normalization of the dark matter signal by a factor $\sim 1.6$ compared to the point source approximation, even for very steep profiles;
\item the estimate of the dark matter particle properties is affected by two sources of bias: (i) the background model and (ii) assumptions on the dark matter annihilation mode; specifically, we showed that different backgrounds can lead to both under- and over-estimates of the dark matter mass, and that the inclusion of annihilation modes with hard spectra biases estimates of the mass towards lower values.
\end{itemize}
We introduced and showed applications of DMFIT, a numerical package that, interfaced with any spectral fitting routine, allows one to fit gamma-ray spectra to the emission of generic WIMP models. DMFIT  includes the $e^+e^-$  annihilation mode, relevant e.g. for Kaluza-Klein dark matter, and is able to fit for low neutralino masses. In addition, we provided an overview and a template of all known gamma-ray sources relevant for dark matter searches in the Galactic center region. While we do not claim here that the Galactic center is the most promising site to search for a signal from dark matter, this study indicates that GLAST data from that region can potentially be of tremendous relevance in the quest for dark matter.


  
\section*{Acknowledgments}
We greatly thank Johann Cohen-Tanugi and Eric Nuss for several comments and suggestions on this manuscript. We also acknowledge useful discussions and inputs from other members of the GLAST collaboration, in particular Jan Conrad, Robert Johnson, Aldo Morselli, Troy Porter and others involved in the Dark Matter / New Physics working group. T.E.J. is grateful for support from the Alexander F. Morrison Fellowship, administered through the University of California Observatories and the Regents of the University of California. 



\end{document}